\documentclass[english,aip,letter,superscriptaddress,floatfix,jcp,nourl,nofootinbib,twocolumn]{revtex4}
\usepackage[T1]{fontenc}
\usepackage{upgreek}
\usepackage{graphicx}
\usepackage{ae,aecompl}
\usepackage{color}
\usepackage{pstricks,pst-node}
\usepackage{psfrag}
\usepackage{verbatim}
\usepackage{bm}
\usepackage{pdflscape}
\usepackage{rotating}
\usepackage{stmaryrd}
\usepackage{amsfonts}
\usepackage{amsmath}
\usepackage{amsthm}
\usepackage{amssymb}
\usepackage{bbold}
\usepackage{endnotes}
\usepackage{ifthen}
\usepackage{bm}
\usepackage{setspace}
\usepackage{endnotes}
\usepackage{fancyhdr}
\usepackage{fancybox}
\usepackage{makeidx}
\usepackage{framed}
\usepackage{listings}
\usepackage{alphalph}
\usepackage{refcount}
\definecolor{shadecolor}{rgb}{0,.7,0}
\definecolor{shadecolor}{gray}{0.7}
\definecolor{shadecolor}{gray}{0.95}
\definecolor{ogreen}{rgb}{0,0.8,0}
\definecolor{magenta}{rgb}{1,0,1}
\definecolor{brown}{rgb}{0.7,0.4,0.2}
\definecolor{shadecolor}{gray}{0.9}

\newcommand{\del}[1]{{\color{red}#1}}
\newcommand{\new}[1]{{\color{blue}#1}}

\newcommand{\Note}[1]{{\bf \color{red}#1}}

\newcommand{\llangle}{\left\langle}
\newcommand{\rrangle}{\right\rangle}
\newcommand{\esc}{\!\cdot\!}

\RequirePackage{dsfont}

\usepackage{xr-hyper}
\usepackage{hyperref}

\graphicspath{{./Figures/}}
\begin{document}

\title{Stochastic Dissipative Euler's equations for a free body}

\author{J.A. de la Torre, J. Sánchez-Rodríguez, Pep Espa\~nol}
\affiliation{ Dept.   F\'{\i}sica Fundamental, Universidad Nacional
  de Educaci\'on a Distancia, Madrid, Spain}

\date{\today}
\begin{abstract}
  Intrinsic  thermal fluctuations  within a  real solid  challenge the
  rigid body assumption  that is central to Euler's  equations for the
  motion of a  free body.  Recently, we have  introduced a dissipative
  and stochastic  version of Euler's equations  in a thermodynamically
  consistent way (European Journal of Mechanics - A/Solids 103, 105184
  (2024)).  This framework describes the evolution of both orientation
  and shape  of a  free body, incorporating  \textit{internal} thermal
  fluctuations and  their concomitant dissipative mechanisms.   In the
  present  work,  we  demonstrate  that, in  the  absence  of  angular
  momentum, the theory predicts that  principal axis unit vectors of a
  body undergo an anisotropic Brownian motion on the unit sphere, with
  the anisotropy arising  from the body's varying  moments of inertia.
  The resulting equilibrium time correlation function of the principal
  eigenvectors decays  exponentially.  This theoretical  prediction is
  confirmed in  molecular dynamics  simulations of small  bodies.  The
  comparison of theory and \textit{equilibrium} MD simulations allow us to measure
  the orientational diffusion tensor.  We then use this information in
  the  Stochastic   Dissipative  Euler's  Equations,  to   describe  a
  \textit{non-equilibrium}  situation of  a body  spinning around  the unstable
  intermediate axis.  The agreement  between theory and simulations is
  excellent, offering a validation of the theoretical framework.

  \end{abstract}
\maketitle
\section{Introduction}

The  motion of  a rigid  body  in free  space is  governed by  Euler's
equations  which  assume that  the  distance  between the  constituent
particles                                                           is
constant\cite{goldsteinClassicalMechanics1983,v.i.arnoldMathematicalMethodsClassical1989}.
This idealization overlooks the inherent  elasticity of solids as well
as the thermal  motion that all particles within a  body experience at
finite  temperatures.   While Euler's  equations  have  been used  for
centuries  to  describe  rotational   dynamics,  they  fall  short  in
explaining  certain  dissipative  phenomena observed  in  nature  like
precession relaxation  or tidal locking. Under  precession relaxation,
free spinning bodies  always end up rotating around the  major axis of
inertia,  explaining  why  98\%  of asteroids  are  spinning  in  pure
rotation   \cite{Lamy1972,Warner2009,Breiter2012}.     Tidal   locking
describes how a gravitationally bounded  body ends up sincronizing its
orbital and  rotational periods, as in  the case of our  Moon.  From a
conceptual point of view, Euler's  equations can be obtained under the
rigid body  idealization from the relationship  between two orthogonal
reference systems in  motion, the observation that  the inertia tensor
diagonalizes  in   the  principal  axis  reference   system,  and  the
conservation   laws,   in   particular  that   of   angular   momentum
\cite{Descamps2008,Gautschi2008}.  However,  until recently,  the link
between the Hamiltonian motion of  the particles constituting the body
and Euler's  equations for the motion  of the body was  lacking.  This
missing  link  has been  addressed  in  Ref.  \cite{espanol2024}.   To
``derive'' Euler's  equations from  Hamilton's equations one  needs to
resort to  the powerful  machinery of  the Theory  of Coarse-Graining,
also known  as Non-Equilibrium Statistical Mechanics  and, some times,
as      the      Mori-Zwanzig      projector      operator      method
\cite{Einstein1905,Green1952,Zwanzig1961,Grabert1982,Ottinger2005}.
In           this           theory,          following           Gibbs
\cite{gibbsElementaryPrinciplesStatistical1960}  one   introduces  the
notions  of   microstates  and   macrostates.   The   macrostates,  or
coarse-grained  (CG) variables,  define  the  coarse-grained level  of
description of the system.  The theory makes the fundamental modelling
assumption that the selected CG  variables evolve in two distinct time
scales,  one  characterized by  small  and  fast contribution  due  to
collisions/vibrations which  is modelled  as white noise,  and another
slow  contribution  due  to  the cummulative  effect  of  these  rapid
contributions.  In this way, the theory describes the evolution of the
CG      variables     as      a      diffusive     Markov      process
\cite{Einstein1905,Green1952}.  One possible  level of description for
the  motion  of a  quasi-rigid  body  would  consider  the body  as  a
continuum,  and  use  the  displacement  and  velocity  fields  as  CG
variables, to produce a viscoelastic field description (including free
boundary conditions) of the free  body.  In the present work, however,
we consider  a coarser  level of description,  closer to  the original
Euler's description,  which takes the  gyration tensor as  the primary
variable.   The gyration  tensor  is closely  related  to the  inertia
tensor and its eigenvectors (defining the principal axis) describe the
orientation of the body, while the eigenvalues (referred to as central
moments) describe the overall shape of the body.

In Ref.  \cite{espanol2024}, we have constructed from first principles
the stochastic differential equations that govern the evolution of the
orientation  and  shape  of  a  quasi-rigid  body.   These  equations,
referred  to  as  Stochastic  Dissipative  Euler's  Equations  (SDEE),
generalize Euler's equations in  order to include thermal fluctuations
and  its  associated  dissipation.    The  reversible  part  of  these
equations is  given by  the usual Euler's  equations, while  the newly
derived  dissipative part  of  the dynamics  contains two  dissipative
mechanisms: an  orientational diffusion of  the principal axis,  and a
dilational friction  damping the oscillations in  the central moments.
In addition,  the SDE describe  the random motion of  the eigenvectors
and eigenvalues due to thermal fluctuations.

For     macroscopic     bodies,    thermal     fluctuations     become
negligible. However,  dissipation cannot be neglected  in general.  In
Ref. \cite{delatorre2024},  we have  studied numerically  the ordinary
differential   equations   that   result   from   neglecting   thermal
fluctuations  in the  SDEE.  By  switching  on and  off the  different
dissipative mechanisms  involved, we  gave support  to the  claim that
precession relaxation  is due  to orientational diffusion  rather than
dilational  friction.  In  fact,  Euler's equations  predict that  the
motion of a  body spinning around the intermediate axis  of inertia is
unstable
\cite{poinsot2022outlines,landauMechanicsThirdEdition1960,Ashbaugh1991},
a result known  as the tennis racket theorem or  the intermediate axis
theorem. As  a consequence,  a free  body experiences  the Dzhanibekov
effect, a striking  phenomena in which the  intermediate axis performs
180$^0$ degrees flips in a  periodic way. When dissipation is included
in the description, the Dzhanibekov effect dissapears as a consequence
of precession  relaxation.  In  the process,  ``organized'' rotational
kinetic energy  is transformed into ``disorganized''  internal thermal
energy while  increasing the entropy  of the system,  and consequently
heating the  body.  The resulting  minimum kinetic energy  occurs when
the body  is spinning  around the  axis of  larger moment  of inertia.

\begin{figure}[t]
  \includegraphics[width=\columnwidth]{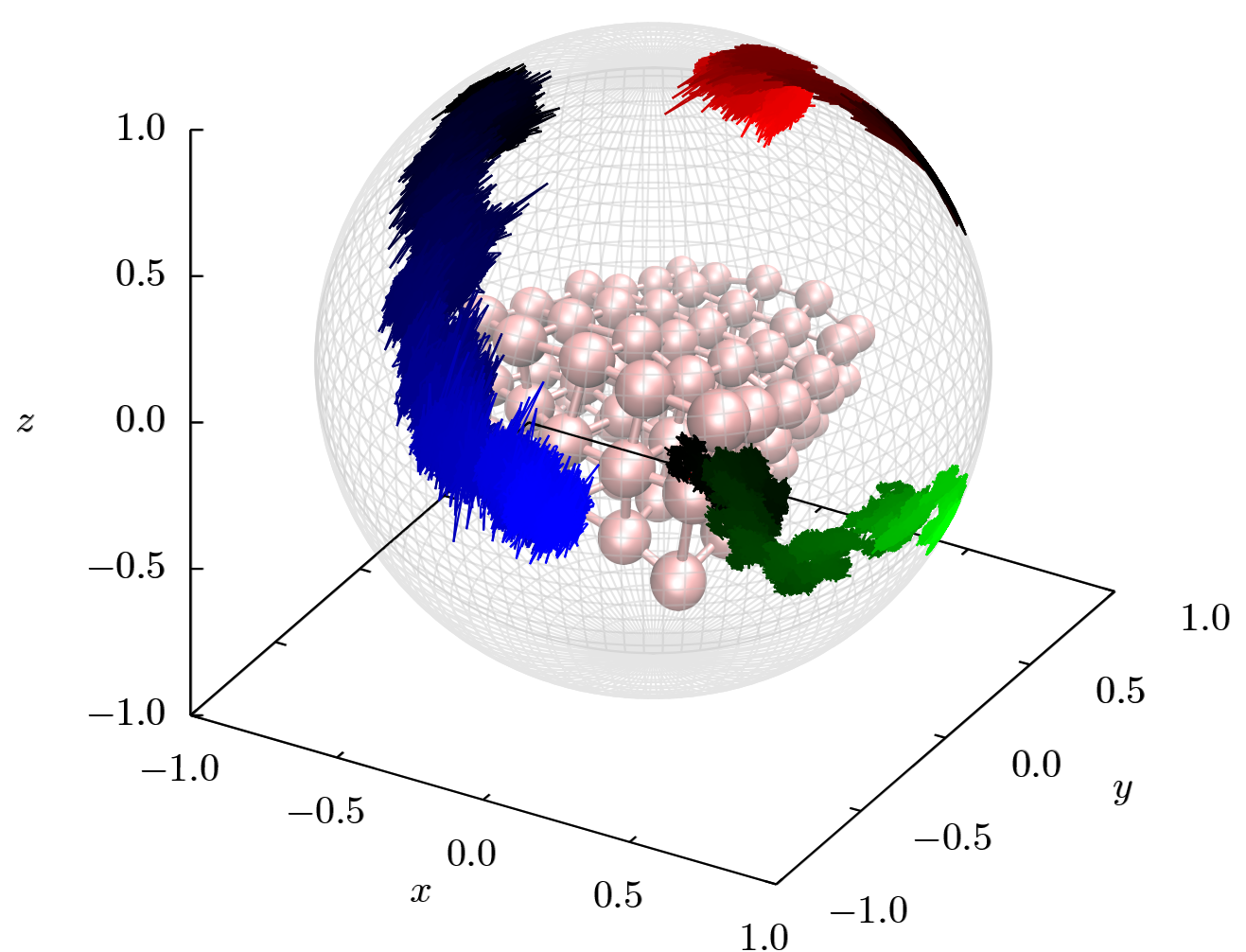}
  \caption{A  small   crystal  of  size  $6\times5\times3$   atoms  is
    simulated with MD. The traces  of the three unit principal vectors
    (in red, green, blue) have  darker colors corresponding to earlier
    times.   We  demonstrate that  these  traces  can be  modelled  as
    realizations  of anisotropic  Brownian motion  on the  unit sphere
    surface. }
  \label{Fig:body}
\end{figure}

In  the  present  paper,  we  turn our  attention  to  the  Stochastic
Dissipative  Euler's  equations   including  thermal  fluctuations  as
formulated in  \cite{espanol2024}.  The objective  of the paper  is to
compare  the  predictions  of  this  theory  with  Molecular  Dynamics
simulations of a  free body composed of bonded  interacting atoms, see
Fig. \ref{Fig:body}. This  constitutes a stringent test  of the theory
and  consitutes a  necessary  validation step.   The SDEE  encompasses
several  parameters  that  must  be  determined  to  enable  numerical
solutions  and generate  predictions.   The list  of these  parameters
include the  average central moments giving  the size and geometry of the  body, an
elastic  matrix defined  in terms  of  the covariance  of the  central
moments, and  two sets  of dissipative coefficients:  an orientational
diffusion matrix and a dilational friction matrix. The values of these
parameters are  obtained from  \textit{equilibrium} MD  simulations of
the body at zero angular momentum.  The test phase of the procedure is
to   predict   \textit{non-equilibrium}   results   about   precession
relaxation using these previously determined set of parameters.

In this paper,  we further elucidate a  theoretical insight concerning
the SDEE derived in \cite{espanol2024}.   We demonstrate that when the
angular momentum  of the body  vanishes, the principal axis  follow an
anisotropic Brownian motion  on the unit sphere.  In this  way, a body
of nanoscopic dimensions  with zero angular momentum  will explore all
possible  orientations \cite{SaportaKatz2019}.   Loosely speaking  the
body ``spins'' without  angular momentum. The key  observation is that
the  movement of  the principal  axes aligns  with the  mathematically
precise      concept      of       spherical      Brownian      motion
\cite{price1983,vandenberg1985}.  The  anisotropic behavior is  due to
the distinct moments of inertia, which introduce a directional bias in
the Brownian  motion.  This  bias reflects  the principle  that larger
moments  of inertia  result  in slower  rotational  motion around  the
corresponding  axis.  We  stress  that  the  physical  origin  of  the
rotational Brownian  motion of  a free body  is the  intrinsic thermal
fluctuations of the particles that  constitute it.  This is physically
different from the usual Brownian  Rotor, where a particle immersed in
a  fluid experiences  Brownian motion  due to  the bombardment  of the
surrounding                                                  molecules
\cite{perrin1934,furry1957,favro1960,hubbard1972,hofling2024}.

As the motion of the principal axis is a spherical Brownian motion, we
predict that  the time-correlation  function of the  principal vectors
decays in a matrix exponential  form.  We compare this prediction with
the  MD   simulation  results,   finding  excellent   agreement.   The
exponential matrix contains  the orientational diffusion coefficients,
and  the  fitting  against  the  MD simulation  results  allow  us  to
determine these parameters.

The  paper is  organized  as follows.   In  Sec.  \ref{Sec:Theory}  we
summarize  the main  result of  Ref. \cite{espanol2024}  which is  the
SDEE. In  Sec \ref{Sec:SDE-S0} we  consider the SDEE when  the angular
momentum  of the  system is  zero,  and show  that the  motion of  the
principal  axis correspond  to  an anisotropic  Brownian motion.   The
details of the mathematical description of anisotropic Brownian motion
are given  in Appendix \ref{Sec:Spherical}.  In  Sec.  \ref{Sec:Eq-MD}
we run equilibrium MD simulations of a small body, and measure all the
required  parameters  in  the  SDEE.  In  Sec.   \ref{Sec:Neq-MD},  we
compare  the numerical  predictions of  the mesoscopic  SDEE with  the
results of  non-equilibrium MD  simulations in which  the body  is set
into  motion through  an ``angular  kick'' that  suddenly sets  a body
initially  at rest  into  rotation along  the  intermediate axis.   We
compare  the  evolution of  the  rotation  kinetic energy,  reflecting
precession relaxation, in the  mesoscopic and microscopic simulations,
obtaining excellent agreement.  Finally, in Sec. \ref{Sec:Conclusions}
we present our conclusions.
\newpage
\section{The motion of a quasi-rigid free body}
\label{Sec:Theory}
In  this   section,  we  summarize   the  theory  presented   in  Ref.
\cite{espanol2024} for  the motion  of a  quasi-rigid free  body. This
allow us to set up the notation used in the paper. The body, which is
composed  of particles  bonded with  a potential,  is described  at a
coarse-grained level  with the center  of mass position  $\hat{\bf R}$
and  gyration tensor  $\hat{\bf G}$,  closely related  to the  inertia
tensor $\hat{\bf I}$.  These CG variables capture how the particles of
the body distribute in space, the first giving the ``location'' of the
body and the  latter giving a sense of its  ``shape and orientation''.
These  variables  are the  ones  used  to  describe  a rigid  body  in
Classical Mechanics under  the rigid constraint assumption,  and it is
natural  to us  these phase  functions as  CG variables.  The gyration
tensor is defined as
\begin{align}
  \label{eq:311}
  \hat{\bf G}&\equiv  \frac{1}{4}\sum_{i}m_{i}\left({\bf r}_{i}-\hat{\bf R}\right)\left({\bf r}_{i}-\hat{\bf R}\right)^T
\end{align}
where the superscript $T$ denotes the transpose matrix.
The prefactor  $1/4$ in  the definition  (\ref{eq:311}) allowed  us in
\cite{espanol2024}    to    interpret   directly    the    eigenvalues
$\hat{  M}_\alpha$  of  $\hat{\bf  G}$ as  ``dilational  masses'',  or
inertia to dilations.  The inertia tensor $\hat{\bf I}$ is defined as
\begin{align}
  \label{InertiaTensor}
  \hat {\bf I} &\equiv  \sum^{N}_{i}m_{i}
[{\bf r}_{i}-\hat{\bf R}]_\times^T
\esc[{\bf r}_{i}-\hat{\bf R}]_\times
\end{align}
where $[{\bf  u}]_\times$ is  the cross-product  matrix formed  from a
vector  ${\bf u}$.   The  action of  the cross  product  matrix on  an
arbitrary vector  ${\bf v}$  gives the cross  product of  both vectors
$ [{\bf u}]_\times\esc {\bf v}={\bf u}\times{\bf v}$.

The  tensors
$\hat{\bf G}$ and $\hat{\bf I}$ are both symmetric, positive definite,
and  they  commute  with  each  other.   As  a  result,  they  can  be
simultaneously  diagonalized   in  the  same  reference   system.  The
principal  axis system,  denoted as  ${\cal S}^0$,  is defined  as the
reference system with its origin at  the center of mass, in which both
tensors   can  be   diagonalized.   Let   $\hat{\bf  e}_\alpha$   with
$\alpha=1,2,3$  be  the  basis  vectors  of  the  inertial  laboratory
reference system ${\cal S}$ and $\hat{\bf e}^0_\alpha$ be the basis of
the  the non-inertial  principal axis  reference system  ${\cal S}^0$.
The components of the rotation matrix  of ${\cal S}^0$ with respect to
${\cal S}$ are defined as
\begin{align}
  \label{eq:142}
    \hat{\boldsymbol{\cal R}}_{\alpha\beta}&=\hat{\bf  e}^T_\beta\esc\hat{\bf  e}^0_\alpha
\end{align}
This shows that $\hat{\bf e}^0_\alpha$  are the  \textit{rows} of the  rotation matrix
$\boldsymbol{\cal R}$.
In ${\cal S}^0$ the gyration and inertia tensor take the form
 \begin{align}
   \hat{\boldsymbol{\cal R}}\esc\hat{\bf G} \esc {\hat{\boldsymbol{\cal R}}}^{T}&= \hat{\mathbb{G}}
                                                                                  \nonumber\\
   \hat{\boldsymbol{\cal R}}\esc\hat{\bf I} \esc {\hat{\boldsymbol{\cal R}}}^{T}&= \hat{\mathbb{I}}
\label{diagonalization}
\end{align}
where $\hat{\mathbb{G}}$ is  a diagonal matrix whose  elements are the
central  moments  $\hat{M}_1,\hat{M}_2,\hat{M}_3$,  which  we  write
compactly           as           a           list           $\hat{{\bf
    M}}=(\hat{M}_1,\hat{M}_2,\hat{M}_3)$. Also,  $\hat{\mathbb{I}}$ is
a   diagonal  matrix   whose  elements   are  the   principal  moments
$\hat{I}_1,\hat{I}_2,\hat{I}_3$.  In this paper, diagonal matrices are
represented with  voided fonts,  as in  $\mathbb{A}$.  As  the inertia
tensor (\ref{InertiaTensor}) can be expressed in terms of the gyration
tensor (\ref{eq:311}) in a linear  way, the eigenvalues of the inertia
tensor are given in terms of the eigenvalues of the gyration tensor as
\begin{align}
  \label{dj}
\hat{I}_\alpha
  &={\black  4}\left(\hat{M}_1+\hat{M}_2+\hat{M}_3-\hat{M}_\alpha\right)
\end{align}
Finally,  observe  that   according  to  (\ref{diagonalization}),  the
vectors $\hat{\bf e}^0_\alpha$ are  the unit eigenvectors of the
gyration and inertia tensors.

The  rotation matrix  can be  expressed  in terms  of the  exponential
matrix
\begin{align}
  \label{eq:206}
\hat{\boldsymbol{\cal R}}=e^{-[\hat{\boldsymbol{\Lambda}}]_\times}
\end{align}
where  $\hat{\boldsymbol{\Lambda}}$ are  the  attitude parameters,  or
orientation    for    short.      The    angular    velocity    vector
$\boldsymbol{\omega}$ is defined as \cite{v.i.arnoldMathematicalMethodsClassical1989,espanol2024}
\begin{align}
\frac{d}{dt}{\boldsymbol{\cal R}}&\equiv-\boldsymbol{\cal R}\esc[\boldsymbol{\omega}]_\times
\label{omega}
\end{align}
The time derivative of the orientation and the angular velocity are related through \cite{espanol2024}
\begin{align}
  \label{eq:29b}  \frac{d{{\boldsymbol{\Lambda}}}}{dt}
  &={\bf B} \esc   {\boldsymbol{\omega}}
\end{align}
where ${\bf B}$ is the Attitude Kinematic Operator \cite{diaz2019} given
by
\begin{align}
  \label{eq:53}
  {\bf B}
  &=\mathbb{1}+{p} [{\bf n}]_\times+{q}[{\bf n}]_\times\esc[{\bf n}]_\times
\end{align}
where        ${\bf        n}=\boldsymbol{\Lambda}/\Lambda$        with
${\Lambda}=|{\boldsymbol{\Lambda}}|$  and  $p,q$   are  the  following
functions of the modulus of the orientation $\Lambda$
\begin{align}
      \label{eq:306}
{p}&=-\frac{{\Lambda}}{2},
   &
{q}=1-\frac{{\Lambda}}{2}\frac{\sin{\Lambda}}{(1-\cos{\Lambda})}
\end{align}

From (\ref{diagonalization}) and (\ref{eq:206}), the gyration tensor can be written as
\begin{align}
  \label{Gdiag}
    \hat{\bf G}&=e^{[\hat{\boldsymbol{\Lambda}}]_\times}\esc
\hat{\mathbb{G}}\esc e^{-[\hat{\boldsymbol{\Lambda}}]_\times}
\end{align}
which can be  understood as a change of variables
from the six independent components  of the symmetric gyration tensor,
and         the         six         degrees         of         freedom
$\hat{\boldsymbol{\Lambda}},\hat{\bf   M}$.   We    have   chosen   in
Ref.  \cite{espanol2024} the  latter as  the primary  CG variables  to
describe the motion of a quasi-rigid free body.

The  CG  variables  are  phase functions,  denoted  with  circumflexed
symbols. As  the microscopic  state of the  body evolves  according to
Hamilton's        equations,         the        phase        functions
$\hat{\boldsymbol{\Lambda}},\hat{\bf  M}$ also  evolve  in time.   The
theory of  CG offers a \textit{modelling}  of the evolution of  the CG
variables  in terms  of  a  diffusive Markov  process.    The corresponding Ito
SDE   for  the   orientation   $\boldsymbol{\Lambda}$   is  given   by
\cite{espanol2024}
\begin{align}
  \label{SDE-Lambda}
  d\boldsymbol{\Lambda}
  &= {\bf B}\esc\left(\boldsymbol{\Omega}
    -\boldsymbol{\Gamma}^{_\Lambda}\esc{\bf B}^{-T}\esc
    (\boldsymbol{\Omega}\times{\bf S})\right)dt
    + k_BT{\bf F}^{\rm th} dt+d\tilde{\boldsymbol{\Lambda}}
\end{align}
where the  spin velocity $  \boldsymbol{\Omega}$ is obtained  from the
conserved  angular  momentum ${\bf  S}$  of  the body  as
\begin{align}
  \boldsymbol{\Omega} &={\bf I}^{-1}\esc{\bf S}
\end{align}
The spin velocity  $ \boldsymbol{\Omega}$ is a  dynamic quantity which
is   different  from   the   angular  velocity   $\boldsymbol{\omega}$
introduced in (\ref{omega}) which is  a purely kinematic quantity. For
a rigid  body both quantities  coincide but for  a real body  they are
different    \cite{espanol2024}.    The    dissipative   matrix    in
(\ref{SDE-Lambda}) is given by
\begin{align}
  \boldsymbol{\Gamma}^{_\Lambda}
  &=    {\bf B}\esc
    e^{[{{\boldsymbol{\Lambda}}}]_\times}\esc
    \boldsymbol{\cal D}_0\esc
     e^{-[{{\boldsymbol{\Lambda}}}]_\times}
    \esc{\bf B}^T
\end{align}
where $\boldsymbol{\cal  D}_0$ is  the orientational  diffusion matrix

The component $\alpha$ of the thermal drift vector $ {\bf F}^{\rm th}$
is given  in terms of  the Kinematic  Operator as (see  Eq. (135) of
\cite{espanol2024})
\begin{align}
  \label{eq:366}
  {\bf F}^{\rm th}_\alpha
  &=    \left(\frac{\partial{\bf B}_{\alpha'\alpha}}{\partial{\Lambda}_\beta}\right)
    \boldsymbol{\cal D}^{\alpha'\beta'}_0{\bf B}_{\beta'\beta}
\end{align}
The noise term in (\ref{SDE-Lambda}) has the form
\begin{align}
  \label{eq:68}
  d\tilde{\boldsymbol{\Lambda}}&=(2k_BT)^{1/2}
    {\bf B}\esc
    e^{[{{\boldsymbol{\Lambda}}}]_\times}\esc
\boldsymbol{\cal D}_0^{1/2}\esc d\tilde{\bf W}
\end{align}
where $d\tilde{\bf W}$  is a vector of independent increments  of the Wiener
process satisfying the mnemotechnical Ito rule
\begin{align}
  \label{eq:293}
  d\tilde{\bf W}d\tilde{\bf W}^T=\mathbb{1}dt
\end{align}
and the square root matrix $\boldsymbol{\cal D}_0^{1/2}$ satisfies
\begin{align}
  \label{eq:468}
  \boldsymbol{\cal D}_0^{1/2}\esc\left(\boldsymbol{\cal D}_0^{1/2}\right)^T&=\boldsymbol{\cal D}_0
\end{align}
Now,  all the  symbols appearing  in  the SDE  (\ref{SDE-Lambda}) for  the
orientation are defined.

The SDEs for the central moments obtained in \cite{espanol2024} are
\begin{align}
  \label{SDE-M}
    d{{\bf M}}&=\boldsymbol{\Pi}dt
              \nonumber\\
  d\boldsymbol{\Pi}&={\boldsymbol{\cal K}}dt-\boldsymbol{\Gamma}\esc\boldsymbol{\nu}dt
                     +   d\tilde{\boldsymbol{\Pi}}
\end{align}
Here $\boldsymbol{\Pi}$ is the  dilational momentum, defined by the first equation
as the time derivative of  the central moments. The $\alpha$ component
of the dilational force is defined as
\begin{align}
  \label{eq:137}
  \boldsymbol{\cal K}_\alpha
  &={\bf M}_{\underline{\alpha}}\left( \frac{1}{2}\boldsymbol{\nu}_{\underline{\alpha}}^2
    +2\left(    \boldsymbol{\Omega}_p^T\esc\boldsymbol{\Omega}_p -\boldsymbol{\Omega}_{p\underline\alpha}^2 \right)\right.
    \nonumber\\
&\quad\left . +\frac{k_BT}{2{\bf M}_{\underline{\alpha}}}
- [\boldsymbol{\Sigma}^{-1}]_{\underline{\alpha}\beta}({\bf M}_\beta-{\bf M}_\beta^{\rm rest})\right )
\end{align}
Here,  repeated   indices  are   summed  over   except  if   they  are
underlined. The need of  breaking this Einstein's summation convention
arises from  the fact that the  central moments do not  transform as a
vector.           The          dilational         velocity          is
$\boldsymbol{\nu}_\alpha=\boldsymbol{\Pi}_\alpha/{\bf  M}_\alpha$ and the
spin     velocity    in     the     principal     axis    frame     is
${\boldsymbol{\Omega}_p}=e^{-[\boldsymbol{\Lambda}]_\times}\esc\boldsymbol{\Omega}$.
The  elasticity matrix  $\boldsymbol{\Sigma}$ is  proportional to  the
covariance of central moments fluctuations, given by
\begin{align}
  \label{eq:138a}
  \boldsymbol{\Sigma}&=\frac{1}{k_BT}\llangle  (\hat{\bf M}-{\bf M}^{\rm rest})(\hat{\bf M}-{\bf M}^{\rm rest})^T\rrangle^{\cal E}
\end{align}
where $\llangle\cdots\rrangle^{\cal  E}$ is  an average with  the rest
microcanonical ensemble  \cite{espanol2024}.  ${\bf M}^{\rm  rest}$ is
the  average  of the  central  moments  with the  rest  microcanonical
ensemble.   We refer  to the  contribution to  ${\boldsymbol{\cal K}}$
quadratic in $\boldsymbol{\nu}$ as the convective term, whose physical
meaning  has been  discussed in  \cite{espanol2024}. The  contribution
quadratic in  $\boldsymbol{\Omega}$ is referred to  as the centrifugal
term,   and   the   last   term  involving   the   elasticity   matrix
$\boldsymbol{\Sigma}$  as   the  elastic   term.   Observe   that  the
dilational momentum  equation in (\ref{SDE-M}) involves  a dissipative
force   $-\boldsymbol{\Gamma}\cdot\boldsymbol{\nu}$  that   is  to   be
interpreted as a dilational friction, where $\boldsymbol{\Gamma}$ is a
dilational friction matrix.

The   temperature   of   the   body   $T$   appearing   in   the   SDE
(\ref{SDE-Lambda}) and (\ref{SDE-M}) depends in general on the thermal
energy ${\cal E}$ of the body. For the solid bodies considered in this work,
typically
\begin{align}
T=  \frac{{\cal E}}{C}
\end{align}
where the  heat capacity is  given by the Dulong-Petit  law $C=3Nk_B$,
where $N$  is the number  of atoms  of the body.   The thermal  energy is
defined as
\begin{align}
{\cal E} &=E-K^{\rm rot}-K^{\rm dil}
\end{align}
where $E$  is the total conserved  energy of the body,  the rotational
kinetic energy is
\begin{align}
  K^{\rm rot}&=\frac{1}{2}{\bf S}\esc{\bf I}^{-1}\esc{\bf S}
               \label{Krot}
\end{align}
and the dilational kinetic energy is
\begin{align}
  \label{Kdil}
  K^{\rm dil}&=\frac{1}{2}\boldsymbol{\Pi}\esc{\mathbb{G}}^{-1}\esc\boldsymbol{\Pi}
\end{align}

The   SDE  (\ref{SDE-Lambda})   for  the   orientation  and   the  SDE
(\ref{SDE-M}) for central moments are coupled through different terms.
For   example,  the   spin  velocity   $\boldsymbol{\Omega}$  in   the
orientation  equation (\ref{SDE-Lambda})  contains the  inertia tensor
${\bf  I}$ that  depends on  the  central moments.   Observe that,  in
principle,       the        orientational       diffusion       tensor
$\boldsymbol{\cal D}_0=\boldsymbol{\cal D}_0({\bf M},{\cal E})$ and it
depends on  the instantaneous value  of the central moments  ${\bf M}$
and the  thermal energy  ${\cal E}$  \cite{espanol2024}. On  the other
hand the equation of the  central moments (\ref{SDE-M}) depends on the
orientation  through  the  centrifugal  term  quadratic  in  the  spin
velocity.

\newpage
\section{The SDE at zero angular momentum}
\label{Sec:SDE-S0}
In this section,  we show that when the angular  momentum of the body
vanishes,  the  theory  in  Ref \cite{espanol2024}  outlined  in  Sec.
\ref{Sec:Theory} characterizes the evolution  of the principal axis as
an anisotropic Brownian motion on the unit sphere.  We offer a detailed description of
anisotropic   Brownian  motion   on   the  sphere   in  the   Appendix
\ref{Sec:Spherical}.

\subsection{The evolution of orientation}
When the  angular momentum  vanishes, ${\bf  S}=0$, (\ref{SDE-Lambda})
with (\ref{BRB}), reduces to the following Ito SDE
\begin{align}
  \label{SDELambda}
  d\boldsymbol{\Lambda}
&=k_BT{\bf F}^{\rm th}dt +     {\bf B}^T\esc{\bf C}\esc d\tilde{\bf W}
\end{align}
where 
\begin{align}
  \label{CD0}
  {\bf C}&\equiv  (2k_BT)^{1/2}\boldsymbol{\cal D}_0^{1/2}
\end{align}
We wish  to obtain the  SDE governing  the evolution of  the principal
axis or, equivalently, of the rotation matrix.  This requires deriving
an                             equation                            for
${\boldsymbol{\cal  R}}=e^{-[{\boldsymbol{\Lambda}}]_\times}$  through
Ito Calculus.  Calculations vastly  simplify by using the Stratonovich
SDE corresponding  to (\ref{SDELambda}) and use  of ordinary calculus.
In   Appendix  \ref{App:2}   we   show  that   the  Stratonovich   SDE
corresponding to the Ito SDE (\ref{SDELambda}) is
\begin{align}
    d\boldsymbol{\Lambda}
&=     {\bf B}^T\esc{\bf C} \circ d\tilde{\bf W}
  \label{SDELambda-S}
\end{align}
From  (\ref{SDELambda-S}) and  by  using the  chain  rule in  ordinary
calculus, we  show in Appendix  \ref{App:3} that the SDE  governing the
rotation matrix is
\begin{align}
  d\boldsymbol{\cal R}
    &=     -\left[{\bf C} \circ d{\bf W}_t\right]_\times\esc \boldsymbol{\cal R}
      \label{R-sphere2}
\end{align}
Therefore, the column vectors  of the rotation matrix satisfy
the Stratonovich SDE
\begin{align}
  d{\bf c}_\alpha
  &= -\left[{\bf C} \circ d{\bf W}_t\right]_\times\esc {\bf c}_\alpha
\end{align}
or in terms of the vector product
\begin{align}
        \label{SDE-c0}
  d{\bf c}_\alpha    ={\bf c}_\alpha\times \left({\bf C}\circ d{\bf W}_t\right)
\end{align}
This  equation  is  identical   to  (\ref{SDE-Str})  in  the  Appendix
\ref{Sec:Spherical} that describes the  anisotropic Brownian motion of
a particle moving on the surface of a unit sphere.  Therefore, the SDE
(\ref{SDELambda}) for the orientation predicts that the unit principal
vectors describe an  anisotropic Brownian motion on the  sphere. It is
straightforward  to show  that  the SDEs  (\ref{SDE-c0}) conserve  the
scalar  products  ${\bf   c}_\alpha\esc{\bf  c}_\beta$  (see  Appendix
\ref{Sec:Spherical}).   Mantaining  these  conservation  laws  with  a
numerical integrator  with finite  time step requires  special methods
\cite{hofling2024}. Instead  of using  (\ref{SDE-c0}), in  the present
paper   we   update    the   orientation   $\boldsymbol{\Lambda}$   in
(\ref{SDELambda})  with  a  predictor-corrector squeme.  This  ensures
automatically that the eigenvalues remain unitary at all times.


\subsection{The evolution of the shape}
Let  us move  now to  the form  of the  SDE (\ref{SDE-M})  for central
moments  when  the  angular  momentum vanishes.   In  this  case,  the
dilational  force  has  no  centrifugal  component  and  (\ref{SDE-M})
reduces to
\begin{align}
  d{\bf M}_\alpha
  &=\boldsymbol{\Pi}_\alpha dt
    \nonumber\\
  d\boldsymbol{\Pi}_\alpha
  &={\bf M}_{\underline{\alpha}}\left( \frac{1}{2}\boldsymbol{\nu}_{\underline{\alpha}}^2
    +\frac{k_BT}{2{\bf M}_{\underline{\alpha}}}
    - [\boldsymbol{\Sigma}^{-1}]_{\underline{\alpha}\beta}({\bf M}_\beta-{\bf M}_\beta^{\rm rest})\right )dt
    \nonumber\\
  &-\boldsymbol{\Gamma}_{\alpha\beta}\boldsymbol{\nu}_\beta dt
    +   d\tilde{\boldsymbol{\Pi}}_\alpha
    \label{SDE-MPi}
\end{align}
Observe that when the body has zero angular momentum, the evolution of
the  central   moments  is  uncoupled   from  the  evolution   of  the
orientation.

A good approximation to the  set of equations (\ref{SDE-MPi}) is given
by the following linearized set of equations
\begin{align}
  \label{Mlinear}
  d{\bf M}_\alpha
  &=\boldsymbol{\Pi}_\alpha dt
    \nonumber\\
  d\boldsymbol{\Pi}_\alpha
  &\simeq -{\bf M}^{\rm eq}_{\underline{\alpha}}[\boldsymbol{\Sigma}^{-1}]_{\underline{\alpha}\beta}({\bf M}_\beta-{\bf M}_\beta^{\rm rest})dt
    \nonumber\\
  &-\boldsymbol{\Gamma}'_{\alpha\beta}\boldsymbol{\Pi}_\beta dt
    +   d\tilde{\boldsymbol{\Pi}}_\alpha
\end{align}
where      we      have      approximated     in      some      places
${\bf M}\simeq{\bf  M}^{\rm eq}$,  have neglected the  convective term
quadratic in  the dilational velocity $\boldsymbol{\nu}$  and the small term
$k_BT/{\bf     M}_\alpha$.      Finally,     we     have     redefined
$\boldsymbol{\Gamma}'_{\alpha\beta}
=\boldsymbol{\Gamma}_{\alpha\underline{\beta}}/{\bf            M}^{\rm
  eq}_{\underline{\beta}}$. With these approximations, the evolution of
the central  moments is an  Ornstein-Ulenbeck (OU) process,  for which
the equilibrium time-correlation of  central moments can be explicitly
computed.  The  equilibrium time correlation functions  oscillate with
frequencies determined by  the elasticity matrix $\boldsymbol{\Sigma}$
which  decay in  a time  scale determined  by the  dilational friction
matrix $\boldsymbol{\Gamma}'$.

\section{Equilibrium MD simulations at ${\bf S}=0$}
\label{Sec:Eq-MD}
In  this  section, we  consider  equilibrium MD simulations of a small body directed  to
measure    the   different    parameters    that    enter   the    SDE
(\ref{SDE-Lambda}), (\ref{SDE-M}). These parameters  will then be used
in non-equilibrium simulations in order to validate the theory.

A parallepiped  crystal made of $6\times5\times3$  atoms that interact
with  a  Lennard-Jones potential  (LJ),  non-linear  bonds, and  angle
contributions is  simulated with LAMMPS  (Large-scale Atomic/Molecular
Massively  Parallel  Simulator)\cite{LAMMPS}.    We  take  $\epsilon$,
$\sigma$ and  $m$ (the mass of  one atom) as fundamental  LJ units, so
that $\tau=\sqrt{m\sigma^2/\epsilon}$ is our unit of time.  Details of
the numerical simulations are given in the Supplemental Material.  The
initial microstate is selected by first carrying out an NVT simulation
to equilibrate  the crystal  to a prescribed  temperature $T$  using a
Nosé-Hoover thermostat.  In order to average different simulations, we
uniformly choose  30 equilibrated microstates  and we subject  each of
them to  a transformation leading to  a fixed value $E$  of the energy
and  to a  zero  value of  the angular  momentum  vector.  Then,  each
microscopic configuration evolves in an  NVE ensemble until the system
is fully equilibrated  at the prescribed temperature.  Once  we have a
typical equilibrium state at energy  $E$ and zero angular momentum, we
consider two  types of  simulations, equilibrium  and non-equilibrium.
In  the equilibrium  simulations we  continue in  the NVE  ensemble to
enter a  production phase in  which we measure  different observables.
In  non-equilibrium  simulations,  we  produce an  angular  kick  that
transforms  the   velocities  in   the  initial   typical  equilibrium
microstate  in such  a  way that  the  system begins  to  rotate at  a
particular initial angular velocity  around the intermediate principal
axis.

\begin{figure*}[t]
  \includegraphics[trim={15 0 35 0},clip,width=0.32\textwidth]{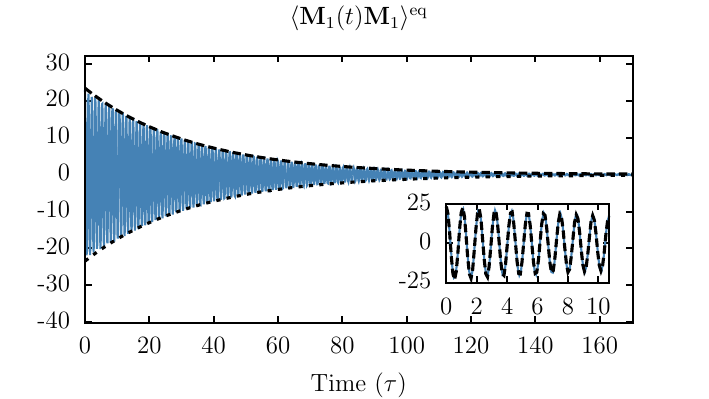}
  \includegraphics[trim={15 0 35 0},clip,width=0.32\textwidth]{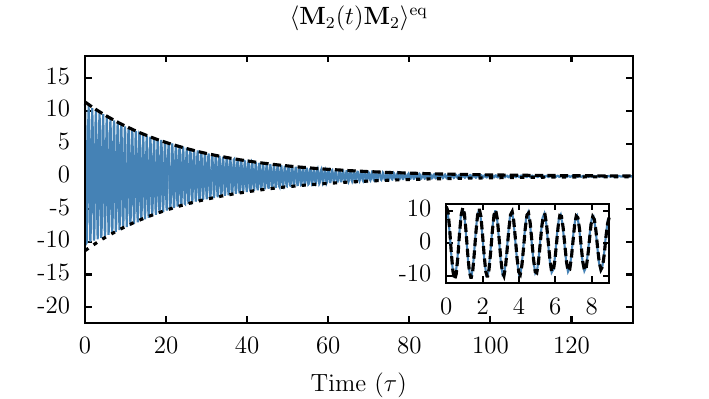}
  \includegraphics[trim={15 0 35 0},clip,width=0.32\textwidth]{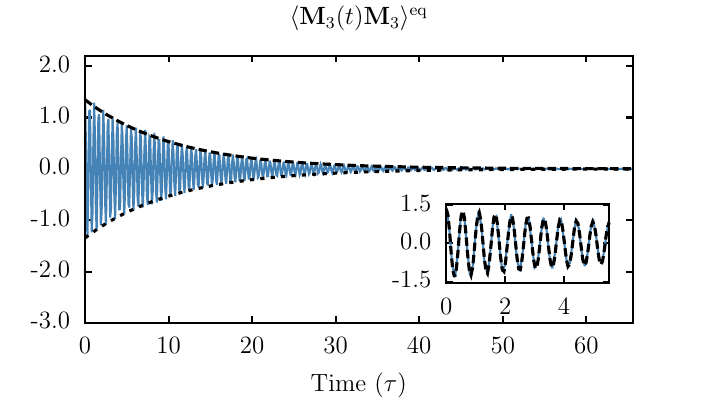}
  \caption{  The equilibrium time correlation
  function          of           the          central          moments
  $\llangle {\bf  M}_\alpha(t) {\bf M}_\alpha\rrangle$ shows  that the
  central moments display a damped oscillatory motion. From the maxima
  and    minima    of   this    curve    we    fit   an    exponential
  $e^{-\Gamma'_\alpha t}$  (in black)  that allows  us to  extract the
  value of the dilational  frictions $\Gamma'_\alpha$. The insets show
  the fitting with (\ref{MtM}).}
  \label{Fig:M}
\end{figure*}
 We set
the total  energy for  the NVE  simulations to  $E=2342\epsilon$. This
value  is a  typical  one  observed in  the  NVT equilibration  phase.
Through the equipartion theorem \cite{espanol2024}
\begin{align}
  \llangle \sum_i\frac{m_i}{2}{\bf v}_i^2\rrangle^{\cal E}_{\rm rest}    =\frac{3(N-2)k_BT}{2}
\end{align}
this corresponds to a temperature of $T=8.7 \epsilon/k_B$. We have checked
that, to  a very  good approximation  the temperature  scales linearly
with the  total energy, $E=C T$  with the heat capacity  following the
Dulong and Petit law $C=3Nk_B$.

The gyration tensor $\hat{\bf G}(t)$ of the body as a function of time is
computed  from (\ref{eq:311}).   The eigenvalues  provide the  central
moments   $\hat{\bf   M}_\alpha(t)$    and   the   unit   eigenvectors
$\hat{\bf e}^0_\alpha(t)$ give the direction of the principal axis.

The  measured equilibrium  averages  of the  central  moments at  rest
${\bf M}^{\rm  rest}$ and the elasticity  matrix $\boldsymbol{\Sigma}$
of the central moments (\ref{eq:138a}) are
\begin{align}
{\bf  M}^{\rm  rest}&=(91.2,62.5,21.0)  m  \sigma^2
  \nonumber\\
  \boldsymbol{\Sigma}&=\left(
                       \begin{array}{ccc}
                         2.635 & 0.003 & 0.001 \\
                         0.003 & 1.273 & 0.001 \\
                         0.001 & 0.001 & 0.162 
                       \end{array}
                       \right) \tau^4 \epsilon
                       \label{ElastMatrix}
\end{align}
We have checked that the  elasticity matrix is practically independent
on the temperature in the range $T=4-8$ in LJ units.  Observe that the
elasticity  matrix  is very  approximately  diagonal.  This entails  a
simplification   in    the   dynamics    of   the    central   moments
(\ref{Mlinear}).  By  assuming  that the  dilational  friction  matrix
$\boldsymbol{\Gamma}'$ is  also diagonal,  the evolution  equations of
the central  moments (\ref{Mlinear})  take the form  (repeated indices
are not summed over here)
\begin{align}
  \label{Mlinear2}
  d{\bf M}_\alpha
  &=\boldsymbol{\Pi}_\alpha dt
    \nonumber\\
  d\boldsymbol{\Pi}_\alpha
  &\simeq -\omega_\alpha^2({\bf M}_{\alpha}-{\bf M}_{\alpha}^{\rm rest})dt
-\Gamma_\alpha'\boldsymbol{\Pi}_{\alpha}dt
    +   d\tilde{\boldsymbol{\Pi}}_\alpha
\end{align}
where
$\omega_\alpha^2={\bf                                          M}^{\rm
  eq}_{\alpha}[\boldsymbol{\Sigma}^{-1}]_{\alpha\alpha}$    are    the
frequencies of oscillation. Using  the values (\ref{ElastMatrix}), the
theoretical frequencies are
\begin{align}
  \label{omega-alpha-theo}
\omega_\alpha\to(5.873, 6.998, 11.363) \tau^{-1}  
\end{align}
    The     equilibrium
time-correlation   for   central   moments,  as   predicted   by   the
Ornstein-Ulhenbeck process is
\begin{align}
  \label{MtM}
  \llangle{\bf M}_\alpha(t) {\bf M}_\alpha\rrangle^{\rm eq}
  &=  \llangle{\bf M}_\alpha {\bf M}_\alpha\rrangle^{\rm eq}
    e^{-\Gamma_\alpha' t}\cos(\omega_\alpha t)
\end{align}
In Fig. \ref{Fig:M} we show  the equilibrium time correlation functions
for  the  central  moments.   To  these curves,  we  have  fitted  the
expressions     (\ref{MtM})     with    the     fitting     parameters
\begin{align}
  \omega_\alpha&\to(5.870, 7.008, 11.357)\tau^{-1}
  \nonumber\\
  \boldsymbol{\Gamma}'
  &=\left(
    \begin{array}{ccc}
0.0295     &0&0
      \\
      0&0.0393  &0
      \\
      0&0&0.0930
    \end{array}\right) \tau^{-1}
    \label{GammaMatrix}
\end{align}
The  agreement  of  the  fitted values  for  $\omega_\alpha$  and  the
predicted values (\ref{omega-alpha-theo}) is very good, and shows that
the modelling  of the  central moments  with a  simple O-U  process is
quite accurate and allows us to measure the dilational friction matrix
$\boldsymbol{\Gamma}'$.

Let us move now to the dynamics  of the unit eigenvectors. As shown in
Fig \ref{Fig:body}, the unit eigenvectors trace a random trajectory as
a  consequence   of  thermal   fluctuations.   We  claim   that  these
MD trajectories  can be modelled  as realizations  of a
Brownian motion on  the unit sphere. To confirm that  this is the case
and  that,  consequently,  the  theory presented  describes  well  the
observed microscopic dynamics of  the orientation  of  the body,  we compute  the
equilibrium time correlation matrix of the eigenvectors
\begin{align}
{\bf E}(t)&\equiv \llangle  \hat{\bf  e}^0(t)\hat{\bf  e}^{0T}\rrangle^{\rm  eq}
\end{align}
As  shown in  the Appendix  \ref{Sec:Spherical}, a  crucial analytical
prediction  of the  SDE (\ref{SDE-c0})  is that  the equilibrium  time
correlation matrix of the eigenvectors is
\begin{align}
  \label{Et}
  {\bf E}(t)
  &=\frac{1}{3}e^{-{\bf A}t}
\end{align}
where  $e^{-{\bf  A}t}$ is  the  exponential  matrix, and  the  matrix
${\bf A}$ is given by
\begin{align}
  \label{DAD0}
  {\bf A}&\equiv  \frac{1}{2}\left({\rm Tr}[{\bf D}]\mathbb{1} -{\bf D}\right)
\end{align}
Here, the matrix ${\bf D}$ is related to the noise amplitude through
\begin{align}
  {\bf D}&\equiv {\bf C}\esc{\bf C}^T\overset{(\ref{CD0})}{=}
           2 k_BT\boldsymbol{\cal D}_0
\end{align}
In terms of $\boldsymbol{\cal D}_0$ the matrix ${\bf A}$ takes the form
\begin{align}
    {\bf A}&
=k_BT
  \left({\rm Tr}[\boldsymbol{\cal D}_0]\mathbb{1} -\boldsymbol{\cal D}_0\right)
             \label{AD0}
\end{align}
while  $\boldsymbol{\cal D}_0$ in terms of ${\bf A}$ is
\begin{align}
  \boldsymbol{\cal D}_0
  &=\frac{1}{k_BT}\left[\frac{1}{2}{\rm Tr}[{\bf A}]\mathbb{1}-{\bf A}\right]
    \label{D0}
\end{align}
as can be easily deduced by taking the trace of (\ref{AD0}).
\begin{figure}[t]
  \includegraphics[width=\columnwidth]{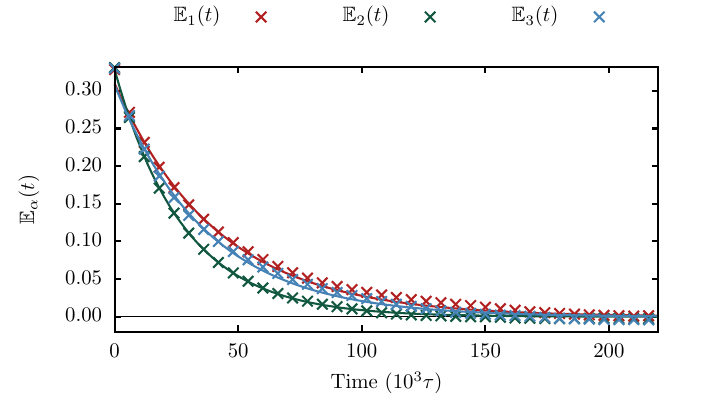} 
  \caption{The   measured    equilibrium   autocorrelation   functions
    $\mathbb{E}_\alpha(t)$ of  the three  principal eigenvectors  as a
    function of time (symbols). Fitted to these curves are exponential
    functions (solid lines), from which the values $\mathbb{A}_\alpha$
    are obtained using (\ref{EA}).}
  \label{Fig:ctc}
\end{figure}
The  measured   off-diagonal  elements   of  the   correlation  matrix
${\bf  E}(t)$  are vanishingly  small.   A  diagonal time  correlation
matrix  ${\bf  E}(t)=\mathbb{E}(t)$  indicates   that  the  matrix  in
(\ref{Et}) is also diagonal  ${\bf A}=\mathbb{A}$. The autocorrelation
function that resides in the diagonal  of ${\bf E}(t)$ is predicted to
decay as a simple exponential function
\begin{align}\label{EA}
  \mathbb{E}_1(t)&=\frac{1}{3}e^{-\mathbb{A}_1t}
                   \nonumber\\
  \mathbb{E}_2(t)&=\frac{1}{3}e^{-\mathbb{A}_2t}
                   \nonumber\\
  \mathbb{E}_3(t)&=\frac{1}{3}e^{-\mathbb{A}_3t}
\end{align}
We  plot in  Fig  \ref{Fig:ctc} the  autocorrelation  function of  the
principal unit  vectors which  decay exponentially, in  full agreement
with  the  theoretical  prediction.    From  the  fitting  values  for
$\mathbb{A}_\alpha$  we  extract  the orientational  diffusion  matrix
$\boldsymbol{\cal D}_0$,  as follows. Because ${\bf  A}=\mathbb{A}$ is
diagonal (\ref{D0}) implies
\begin{align}
    \boldsymbol{\cal D}_0
  &=\frac{1}{k_BT}\left[\frac{1}{2}{\rm Tr}[\mathbb{A}]\mathbb{1}-\mathbb{A}\right]
    \label{D02}
\end{align}
which is the diagonal matrix
\begin{align}
      \boldsymbol{\cal D}_0
  &=\left(
    \begin{array}{ccc}
    \frac{  \mathbb{A}_2+\mathbb{A}_3-\mathbb{A}_1}{2k_BT}      &0&0
      \\
      0&\frac{\mathbb{A}_1+\mathbb{A}_3-\mathbb{A}_2}{2k_BT}  &0
      \\
      0&0&\frac{\mathbb{A}_1+\mathbb{A}_2-\mathbb{A}_3}{2k_BT}
    \end{array}\right)
\end{align}
From the  fitted values  of $\mathbb{A}_\alpha$ in  Fig. \ref{Fig:ctc}
the matrix takes the value
\begin{align}
      \boldsymbol{\cal D}_0
  &=\left(
    \begin{array}{ccc}
      2.2338 & 0 & 0
      \\
      0& 0.8587  &0
      \\
      0&0&1.9129
    \end{array}\right) \times 10^{-6} (\tau \epsilon)^{-1}
    \label{D0Matrix}
\end{align}
In  summary, the  exponential  decay of  the  autocorrelation of  unit
eigenvalues allow us to measure the orientational diffusion matrix.
\section{Non-equilibrium simulations with ${\bf S}\neq 0$}
\label{Sec:Neq-MD}
In  this  section,  we   will  compare  results  from  non-equilibrium
microscopic MD simulations with non-equilibrium mesoscopic simulations
of  the SDEs  (\ref{SDELambda}),(\ref{SDE-MPi}) for  a spinning  body.
The  parameters  to  be  used  in  the  SDEs  have  been  obtained  in
(\ref{ElastMatrix}),(\ref{GammaMatrix}),(\ref{D0Matrix}).  Because the
size  of the  system  is  small (the  crystal  has  90 atoms)  thermal
fluctuations are rather large and the  signal to noise ratio is small.
Therefore, in  order to compare  the noisy signals in  both simulation
methods, and  to be able  to validate the  theory, it is  necessary to
perform some  sort of averaging.  On  one hand, the average  in the MD
simulations will be  over initial microstates compatible  with a given
(non-equilibrium) macrostate, i.e., with the same energy, angular momentum,
and central  moments.  The first  two observables (energy  and angular
momentum) are conserved  by construction of the  algorithm.  To obtain
microscopic configurations with the same value for the central moments
${\bf M}$, we carry out a long simulation in the NVE ensemble, compute
the  central moments  along the  entire simulation,  and select  those
microstates  which have  similar values  of ${\bf  M}$.  This  is done
through  a Kmeans  algorithm  that cluster  the  central moments  into
groups of equal variance, minimizing the within-cluster sum-of-squares
distance to the centroid of each group.  \cite{scikit-learn} Following
this  procedure,  we select  a  cluster  of 50  different  microscopic
configurations  with equal  energy and  angular momentum,  and central
moments differing  by less  than 1\%.  Note that we  do not  group the
microstates   with   similar   values  of   the   dilational   momenta
$\boldsymbol{\Pi}$, which fluctuates  around zero and seems  to have no
big effect on the results.

On the other hand, the average in  the SDE will be done over different
realizations  of  the  stochastic  process starting  from  an  initial
non-equilibrium macrostate  identical to  the one  ocurring in  the MD
simulations.

\subsection{MD simulations}
\begin{figure}[t]  \includegraphics[width=\columnwidth]{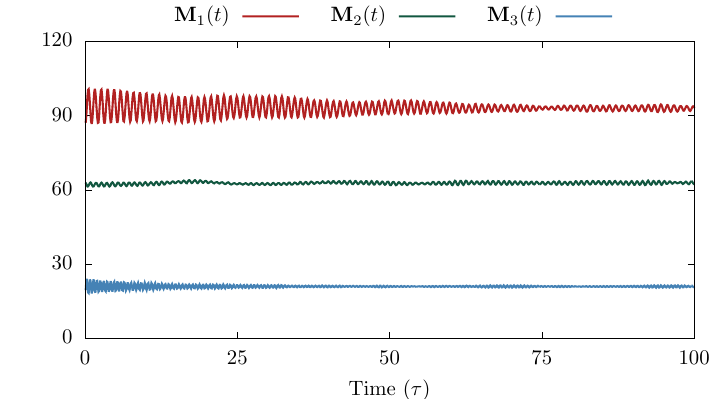}
  \caption{Averaged  central moments  over 50  simulations, after  the
    ``angular  kick''.   From  top   to  bottom:  $M_1$,   $M_2$,  and
    $M_3$. Observe that the central  moment $M_2$ corresponding to the
    intermediate axis,  is hardly affected by  the centrifugal forces,
    while  the decay  time scales  for $M_1$  and $M_3$  is different,
    reflecting different values of the dilational friction.}
  \label{Fig:Mtrot}
\end{figure}
\begin{figure}[t]
  \includegraphics[width=\columnwidth]{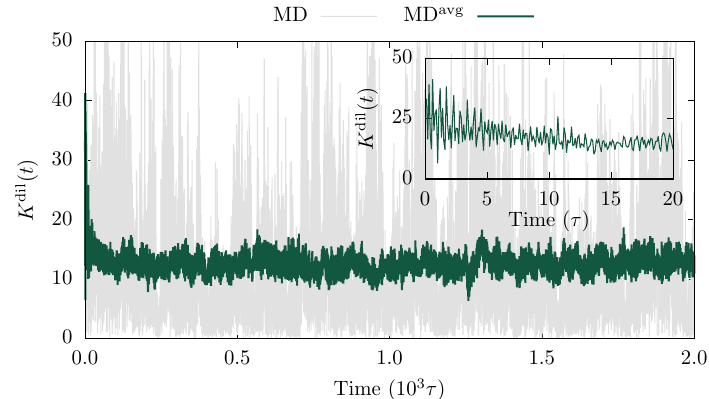}
  \caption{Dilational  energy  $K^{\mathrm{dil}}(t)$  defined in (\ref{Kdil}) for  an  initial
    angular  velocity ${\Omega}  =  0.75 \tau^{-1}$.  Green
    line corresponds to  the averaged value over  50 different initial
    conditions with the same value $E$,  ${\bf S}$ and ${\bf M}$. Gray
    line corresponds  to an individual  MD simulation. Inset:  zoom of
    the initial behavior of $K^{\mathrm{dil}}(t)$ that shows the decay
    of the dilational energy after the angular kick.}
  \label{Fig:Kdil}
\end{figure}
\begin{figure*}[t]
  \includegraphics[width=0.3\textwidth]{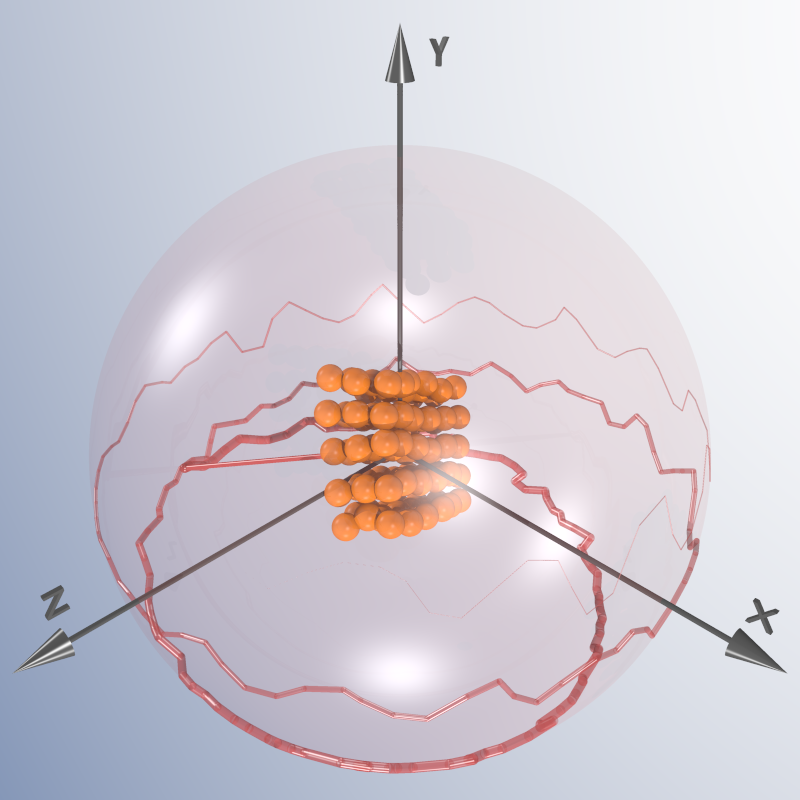}
  \includegraphics[width=0.3\textwidth]{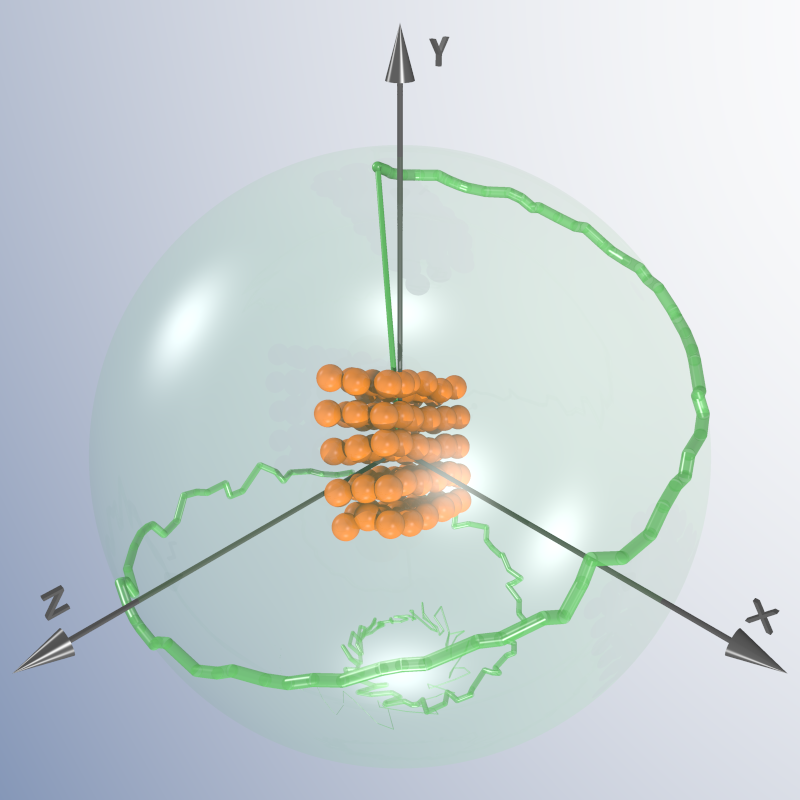}
  \includegraphics[width=0.3\textwidth]{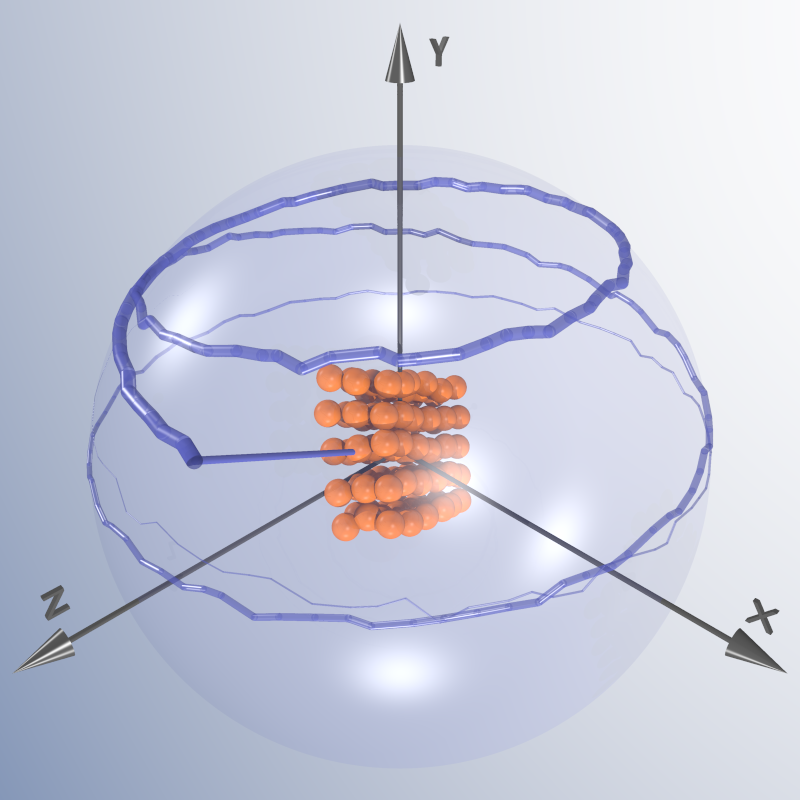}
  \caption{Time evolution  (plotted with increasing thickness)  of the
    principal  vectors  ${\bf  e}_1(t)$ (red,  left),  ${\bf  e}_2(t)$
    (green,  middle),  and  ${\bf   e}_3(t)$  (blue,  right)  for  one
    particular   MD   simulation.    The  Dzhanibekov   effect   where
    ${\bf  e}_2(t)$  flips direction  is  noticeable:  at early  times
    (thiner thickness of the line) ${\bf e}_2(t)$ is pointing upwards,
    at  midle times  it  is  downwards, and  at  later times  (thicker
    thickness) it is pointing upwards again.}
  \label{Fig:Dzhanibekov}
\end{figure*}

\begin{figure}[h]
  \includegraphics[width=\columnwidth]{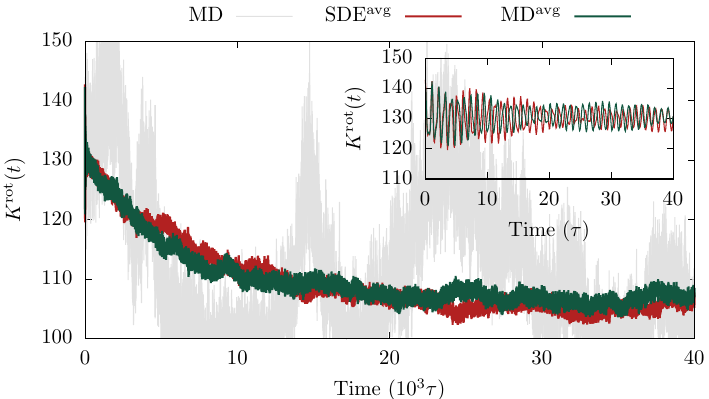}
  \caption{Comparison  of   the  average  rotational   kinetic  energy
    $K^{\rm    rot}(t)$    for    an    initial    angular    velocity
    ${\Omega}=0.75\tau^{-1}$   and  the   same  values   of
    $E,{\bf S},{\bf M}$.   Red line corresponds to an  average over 50
    different  realizations  of  the   SDE  simulations.   Green  line
    correspondos  to  the  average   over  50 initial  conditions  in  MD
    simulations.    Gray  line   corresponds  to   an  individual   MD
    simulation.    Inset:   Zoom   of    the   initial   behavior   of
    $K^{\rm  rot}(t)$ that  captures the  fact that  the angular  kick
    excites  oscillatory   motion  of  the  central   moments  through
    centrifugal forces.   This effect  is also  reproduced in  the SDE
    simulations.  The  time scale of  the initial oscillations  in the
    inset  coincides with  the time  scale of  the damping  of central
    moments at equilibrium, as shown in Figs.  \ref{Fig:body}.}
  \label{Fig:Krot-MD-SDE}
\end{figure}

The  non-equilibrium  MD  simulations  are performed  as  follows.   A
microstate  $z$  is selected  from  the  equilibrium simulations  with
prescribed values  of $E,{\bf  S},{\bf M}$.  Then,  a rotation  of the
crystal is performed to reorient it  with the intermediate axis in the
$y$ axis,  and an ``angular  kick'' is  imparted, as described  in the
Supplemental Material.   This angular  kick transforms  the microstate
$z$       to      $z'$       with       an      angular       velocity
$\boldsymbol{\Omega}=(0,\Omega,0)$ and the  same original total energy
$E$.  We choose a value of  $\Omega=0.75 \tau^{-1}$. If we run now the
MD simulation starting from the  new microstate $z'$, the crystal will
rotate  around   the  intermediate  axis  with   an  angular  momentum
${\bf   S}   =   {\bf   I} \cdot   \boldsymbol{\Omega}   \equiv   {\bf
  S}^{\mathrm{eq}}$.  After the applied  angular kick, the centrifugal
force  produces  an  inital  expansion  of  the  crystal  that  starts
oscillating, producing  an oscillatory  motion of the  central moments
which  is   superimposed  to   their  thermal   motion  as   shown  in
Fig. \ref{Fig:Mtrot}.  Eventually these  oscillations damp out and the
dilational  kinetic energy  $K^{\rm dil}(t)$  defined in  (\ref{Kdil})
goes   down  to   its  thermal   noise   level,  as   shown  in   Fig.
\ref{Fig:Kdil}.    After  the   initial   angular   kick  around   the
intermediate  axis, the  system rotates  and displays  the Dzhanibekov
effect in  very much the same  way as in the  fully deterministic case
\cite{delatorre2024}. This  effect is  appreciable from  the flip-flop
evolution   of    the   intermediate    axis   as   shown    in   Fig.
\ref{Fig:Dzhanibekov}.  At  long times,  the crystal ends  up spinning
around  the  major  axis,  which aligns  with  the  conserved  angular
momentum vector.  To keep track of the evolution of the system towards
the equilibrium  final state, we  choose as observable  the rotational
kinetic energy $\hat{K}^{\rm rot}$  defined in (\ref{Krot}).  A single
realization  of   the  MD   simulation  is  shown   in  gray   in  Fig
\ref{Fig:Krot-MD-SDE}, where it is apparent  that the noise level does
not  allow for  a clear  distinction  between signal  and noise,  thus
justifying the averaging  over 50 different simulations  with the same
$E$, ${\bf S}$ and similar central moments.

\subsection{SDE simulations}
We carried out 50 simulations of the SDEs with the same initial values
of  the orientation,  central moments  and dilational  momenta of  the
corresponding     non-equilibrium    MD     simulation,    using     a
predictor-corrector     explicit      scheme\cite{Delong2014}.      In
Fig.~\ref{Fig:Krot-MD-SDE} we  compare the  average of  the rotational
kinetic energy $K^{\mathrm{rot}}$  obtained from the SDE  (red) and MD
(green) simulations.  The rotational kinetic energy starts and ends at
the same values in both SDE and  MD, as a consequence of the identical
initial  macrostate selected.   The  stochastic precession  relaxation
produces a  spinning about  the major axis.   The initial  damping and
long-term decay  time of  the MD  signal is well  captured by  our SDE
model. The  very good agreement of  the MD and SDE  results provides a
further validation of the proposed theory.

\newpage
\section{Conclusions}
\label{Sec:Conclusions}

In this paper we have compared the predictions  of the
Stochastic           Dissipative           Euler's           Equations
(\ref{SDE-Lambda}),(\ref{SDE-M})   with  the   results  of   Molecular
Dynamics simulation.   Fitting parameters at equilibrium  allows us to
predict  the  non-equilibrium  precession   relaxation  of  the  body,
reflected  in   the  evolution  of  the   average  rotational  kinetic
energy. In addition,  we confirm through the exponential  decay of the
autocorrelation function  of the principal unit  eigenvectors that the
motion of  these vectors  can be modelled  as an  anisotropic Brownian
motion, where  the stochasticity is  intrinsic, rather than due  to an
external bath (gas or liquid).   The agreement between the theoretical
model and the  simulations is excellent, thus  confirming the validity
of the former.

We have observed that orientational  diffusion strongly depends on the
value of the rest central moments.   Changing the size and geometry of
the  body changes  strongly  the value  of  the coefficients.   Larger
bodies display  very small orientation fluctuations.  In addition, the
correlation time  of the principal  axis also depends strongly  on the
body's  dimensions,  increasing  with  size.   As  a  consequence,  MD
simulations are readily  unfeasible to access the long  time scales in
which a  non-spinning body  will change  its overall  orientation.  In
order to estimate the orientational diffusion coefficients from MD, we
can only  consider small bodies.  We are currently attempting  to find
scaling relationship between geometry  and the orientational diffusion
coefficients. This requires extremely  long simulation times, and will
be presented  elsewhere.  On  the other hand,  using too  small bodies
(like molecules) may break one of the implicit assumptions made in the
Theory of Coarse-Graining used to derive the SDEE, which is the mixing
assumption  of  the  Hamiltonian  flow ensuring  the  existence  of  a
well-defined  equilibrium  state.    Not  all  interaction  potentials
between the particles of the  body ensure this property.  For example,
using almost  linear harmonic  springs does not  give, in  general, an
ergodic  system, as  the  famous  Fermi-Pasta-Ulam-Tsingou problem  of
thermalization                                                  showed
\cite{fermi,Berman2005,Dauxois2005,Dauxois2008a}.  We  have dealt with
this problem  by ensuring sufficiently non-linear  interactions, based
on  the Lennard-Jones  potential.  The  model atomic  potential energy
considered in the  present paper does not correspond  necessarily to a
real material.  Future  work should focus on  realistic potentials for
diamond and  silica, for example.  Further  comparisons between theory
and  MD simulations  are strongly  constrained  to a  small window  of
system sizes: smaller sizes may not be ergodic, larger body sizes lead
to  extremely long  correlation times.   It is  possible to  play with
temperature,  as   we  have  observed  that   orientational  diffusion
increases with the temperature of the  body.  In the present paper, in
order to have observable effects that allow for a quantitative precise
estimation  of $d_\alpha$  we have  considered very  high temperatures
($k_BT\simeq9\epsilon$) that  may not be entirely  realistic.  Because
the model selected does not correspond necessarily to a real material,
we are not  yet able to provide accurate estimates  of the time scales
involved  in the  precession  relaxation rate  for  real materials  of
realistic sizes.

The present theory describes  orientational diffusion due to intrinsic
thermal fluctuations.  The phenomena is very different from rotational
diffusion due  to the  interaction with an  ambient gas  that produces
additional friction. This latter phenomenon  should be modelled with a
Brownian          rotor         with          viscous         friction
\cite{perrin1934,furry1957,favro1960,hubbard1972,hofling2024}  or with
kinetic  theory  \cite{martinetz2018a}.   For   a  dense  gas,  it  is
anticipated  that the  friction with  the gas  will dominate  over the
intrinsic orientational  diffusion.  It is  an open question  at which
gas densities the intrinsic rotational diffusion, as discussed in this
paper, will be surpassed by  the effects of gas friction.  Interaction
with radiation  as it is encountered  in optolevitodynamic experiments
is  also a  source  of stochasticity  due to  the  discrete nature  of
photons. The  comparison of the  magnitude of these  different effects
clearly deserves further study.  The  significance of the present work
lies in its  validation of the theoretical model by  comparing it with
Molecular  Dynamics   (MD)  simulations   in  situations   where  such
comparisons are feasible.


\section*{Acknowledgments}
We  thank  Mark  Thachuck  for  his  useful  comments  regarding  this
manuscript.  This  research has  been  supported  through MCIN  grants
PDC2021-121441-C22  and  PID2020-117080RB-C54.  JSR  acknowledges  the
support of  the Ministerio de  Ciencia, Innovación y  Universidades of
Spain  under  a  Margarita  Salas  contract  funded  by  the  European
Union-NextGenerationEU.   We acknowledge  the computational  resources
and  assistance  provided  by  the   Centro  de  Computación  de  Alto
Rendimiento CCAR-UNED.

\section{Appendix: Brownian motion on a sphere}
\label{Sec:Spherical}
In  this  appendix, we  summarize  some  mathematical results  on  the
Brownian  motion of  a particle  on the  surface of  a sphere  and its
anisotropic generalization.

\subsection{Ito vs Stratonovich}
We  first  recall the  connection  between  the Ito  and  Stratonovich
interpretation of an SDE.  The following Ito SDE
\begin{align}
  d{\bf x}={\bf A}({\bf x})dt+\boldsymbol{\Theta}({\bf x})\esc d{\bf W}_t
  \label{SDE-I}
\end{align}
with ${\bf A}({\bf x})$  the drift, $\boldsymbol{\Theta}({\bf x})$ the
noise  amplitude matrix,  and  $d{\bf W}_t$  a  vector of  independent
increments  of  the  Wiener  process,  corresponds  to  the  following
Stratonovich SDE \cite{Gardiner1983}
\begin{align}
    d{\bf x}=\left[{\bf A}({\bf x})-{\bf V}({\bf x})\right]dt+\boldsymbol{\Theta}({\bf x})\circ d{\bf W}_t
  \label{SDE-S}
\end{align}
where the stochastic drift is 
\begin{align}
  {\bf V}_i
  &=\frac{1}{2}\boldsymbol{\Theta}_{kj}\frac{\partial}{\partial {\bf x}_k}\boldsymbol{\Theta}_{ij}
    \label{Vi}
\end{align}
and  repeated  indices are  summed  over.   Both SDE  (\ref{SDE-I}),
(\ref{SDE-S}) correspond to the following Fokker-Planck Equation (FPE)
\begin{align}
  \partial_tP({\bf x},t)
  &=-\frac{\partial}{\partial {\bf x}}{\bf A}P({\bf x},t)
+\frac{1}{2}\frac{\partial}{\partial {\bf x}}\frac{\partial}{\partial {\bf x}}\boldsymbol{\Theta}\boldsymbol{\Theta}^T P({\bf x},t)
    \label{FPE}
\end{align}

\subsection{Anisotropic Brownian motion on the sphere}
It is well-known \cite{price1983,vandenberg1985}  that the evolution of
a particle with position ${\bf  r}\in\mathbb{R}^3$ on the surface of a
sphere due to Brownian motion is described by the following Stochastic
Differential Equation (SDE)
\begin{align}
  \label{SDE-sphere}
  d{\bf r}
  &=[{\bf r}]_\times \circ d{\bf W}_t
  &&\mbox{Stratonovich}
\end{align}
with the corresponding SDE in the Ito interpretation
\begin{align}
  d{\bf r}
  &=-{\bf r}+[{\bf r}]_\times\esc d{\bf W}_t
  &&\mbox{Ito}
\end{align}
where  $d{\bf  W}_t\in\mathbb{R}^3$  is  a  a  vector  of  independent
increments  of  the  Wiener  process and  $[{\bf  r}]_\times$  is  the
cross-product matrix formed from the  vector ${\bf r}$.  

A natural generalization of  the SDE (\ref{SDE-sphere}) for describing
\textit{ non-isotropic} Brownian motion on the sphere is
\begin{align}
  \label{SDE-Str}
  d{\bf r}
  &=[{\bf r}]_\times \esc{\bf C}\circ d{\bf W}_t &&\mbox{Stratonovich}
\end{align}
where ${\bf  C}\in\mathbb{R}^{3\times3}$ is a constant symmetric  matrix.
We write this equation in the form (\ref{SDE-S})
\begin{align}
  d{\bf r}=\boldsymbol{\Theta}({\bf r})\circ d{\bf W}_t
  \label{deStrat}
\end{align}
where
\begin{align}
  \boldsymbol{\Theta}({\bf r})=[{\bf r}]_\times\esc{\bf C}
\end{align}
and ${\bf C}$ is a constant matrix.  The
corresponding Ito SDE  is given by
\begin{align}
  \label{SDEIto}
  d{\bf r}&={\bf V} dt + \boldsymbol{\Theta}\esc d{\bf W}_t
\end{align}
where ${\bf V}$ is given in (\ref{Vi}). Noting that the matrix $\boldsymbol{\Theta}$ has components
\begin{align}
  \boldsymbol{\Theta}_{ij}=\epsilon_{ilm}{\bf r}_l{\bf C}_{mj}
\end{align}
then
\begin{align}
    {\bf V}_i({\bf x})
  &=\frac{1}{2}\boldsymbol{\Theta}_{kj}\frac{\partial}{\partial {\bf r}_k}\boldsymbol{\Theta}_{ij}
    \nonumber\\
  &=\frac{1}{2}\epsilon_{kl'm'}{\bf r}_{l'}{\bf C}_{m'j}\frac{\partial}{\partial {\bf r}_k}\epsilon_{ilm}{\bf r}_l{\bf C}_{mj}
    \nonumber\\
  & =-\frac{1}{2}\left({\rm Tr}[{\bf C}\esc{\bf C}^T]{\bf r}_{i}
    -{\bf C}_{ij}{\bf C}_{mj}{\bf r}_{m}\right)
\end{align}
which means
\begin{align}
  {\bf V}&=  -\frac{1}{2}\left({\rm Tr}[{\bf D}]\mathbb{1} -{\bf D}\right)\esc{\bf r},
\end{align}
where we have defined the symmetric positive definite diffusion tensor 
\begin{align}\label{D}
{\bf D}\equiv{\bf C}\esc{\bf C}^T
\end{align}
and the  superscript $T$
denotes the transpose  matrix. The Ito SDE  (\ref{SDEIto}) is then
\begin{align}
  \label{SDE-Ito}  d{\bf r}
  &=-{\bf A}\esc{\bf r}+
            [{\bf r}]_\times\esc{\bf C}\esc d{\bf W}_t&&\mbox{Ito}
\end{align}
where the matrix ${\bf A}\in\mathbb{R}^{3\times3}$ is
\begin{align}
  \label{A}
{\bf A}\equiv  \frac{1}{2}\left({\rm Tr}[{\bf D}]\mathbb{1} -{\bf D}\right)
\end{align}
Here $\mathbb{1}$ is the unit matrix.  The matrix ${\bf A}$ is symmetric
and positive  definite. This  can be  easily seen  from the  fact that
${\bf A},{\bf D}$ commute and, hence diagonalize in the same basis. If
the eigenvalues of ${\bf D}$ are $(d_1,d_2,d_3)$, then the eigenvalues
of      ${\bf      A}$      are       easily      seen      to      be
$\left(\frac{d_2+d_3}{2},\frac{d_2+d_3}{2},\frac{d_1+d_2}{2}\right)$. Because
$d_\alpha>0$, all  the eigenvalues  of ${\bf  A}$ are  positive.  When
${\bf   C}=\mathbb{1}={\bf  D}$,   then   ${\bf  A}=\mathbb{1}$,   and
(\ref{SDE-Str})  recovers  the  SDE (\ref{SDE-sphere})  for  isotropic
Brownian   motion.

\begin{widetext}
  \subsection{The FPE for anisotropic Brownian motion}
  To get  the FPE that corresponds  to the Ito SDE  (\ref{SDE-Ito}) we
  translate the connection between the FPE (\ref{FPE}) and the Ito SDE
  (\ref{SDE-I}), this is
\begin{align}
  \label{FPEIto}
  \partial_tP({\bf r},t)
  &=\frac{\partial}{\partial {\bf r}}
    \left[\frac{1}{2}\left({\rm Tr}[{\bf D}]\mathbb{1} -{\bf D}\right)\esc{\bf r}\right]
    P({\bf r},t)
    +\frac{1}{2}\frac{\partial}{\partial {\bf r}}
    \frac{\partial}{\partial {\bf r}}[{\bf r}]_\times\esc{\bf C}\esc{\bf C}^T[{\bf r}]^T_\times P({\bf r},t)
    \nonumber\\
  &=\frac{\partial}{\partial {\bf r}}
    \left[\frac{1}{2}\left({\rm Tr}[{\bf D}]\mathbb{1} -{\bf D}\right)\esc{\bf r}\right]
    P({\bf r},t)
    +\frac{1}{2}\frac{\partial}{\partial {\bf r}}
   [{\bf r}]_\times\esc{\bf D}\esc [{\bf r}]^T_\times  \frac{\partial}{\partial {\bf r}}P({\bf r},t)
    +\frac{1}{2}\frac{\partial}{\partial {\bf r}}\left(
    \frac{\partial}{\partial {\bf r}}[{\bf r}]_\times\esc{\bf D}\esc [{\bf r}]^T_\times \right)P({\bf r},t)
\end{align}

Let us compute the term within rounded parenthesis
\begin{align}
  \left(
  \frac{\partial}{\partial {\bf r}_j}\left[[{\bf r}]_\times\esc{\bf D}\esc [{\bf r}]^T_\times \right]_{ij}\right)
  &=-
    \frac{\partial}{\partial {\bf r}_j}\epsilon_{iki'}{\bf D}_{i'j'}\epsilon_{j'k'j}
    \frac{\partial}{\partial {\bf r}_j}{\bf r}_k{\bf r}_{k'}
    =-\epsilon_{iki'}\epsilon_{j'k'j}{\bf D}_{i'j'}\left(\delta_{kj}{\bf r}_{k'}+\delta_{k'j}{\bf r}_k\right)
    \nonumber\\
  & =-\epsilon_{iji'}\epsilon_{j'k'j}{\bf D}_{i'j'}{\bf r}_{k'}-\epsilon_{iki'}\underbrace{\epsilon_{j'jj}}_{=0}{\bf D}_{i'j'}{\bf r}_k
  \nonumber\\
  &=\epsilon_{ii'j}\epsilon_{j'k'j}{\bf D}_{i'j'}{\bf r}_{k'}
    =\left[\delta_{ij'}\delta_{i'k'}-\delta_{ik'}\delta_{i'j'} \right]
    {\bf D}_{i'j'}{\bf r}_{k'}
    =\delta_{ij'}\delta_{i'k'}{\bf D}_{i'j'}{\bf r}_{k'}-\delta_{ik'}\delta_{i'j'}{\bf D}_{i'j'}{\bf r}_{k'}
    \nonumber\\
  & ={\bf D}_{i'i}{\bf r}_{i'}-{\rm Tr}[{\bf D}]{\bf r}_{i}
\end{align}
\end{widetext}
The last term in the rhs of (\ref{FPEIto}) then cancels the first term
and the FPE (\ref{FPEIto}) for anisotropic Brownian motion is simply
\begin{align}
  \label{FPEfin}
  \partial_tP({\bf r},t)
  &=
    \frac{1}{2}\frac{\partial}{\partial {\bf r}}
   [{\bf r}]^T_\times\esc{\bf D}\esc [{\bf r}]_\times  \frac{\partial}{\partial {\bf r}}P({\bf r},t)
\end{align}
The diffusion matrix of this  FPE is clearly positive (semi) definite,
because, for any arbitrary vector ${\bf v}$ we have
\begin{align}
  {\bf v}^T\esc    [{\bf r}]^T_\times\esc{\bf D}\esc [{\bf r}]_\times\esc{\bf v}
  &=  {\bf v}^T\esc    [{\bf r}]^T_\times\esc{\bf C}\esc{\bf C}^T\esc [{\bf r}]_\times\esc{\bf v}
 \nonumber\\
  &    =\left({\bf C}^T\esc [{\bf r}]_\times\esc{\bf v}\right)^2\ge0
\end{align}

\subsection{The equilibrium distribution for spherical Brownian motion}
Let us  find the equilibrium  distribution of the  FPE (\ref{FPEfin}).
Observe that  any probability of  the form $P({\bf  r})=\phi(r)$, with
$r=|{\bf  r}|$, is  a  stationary solution  of  the FPE  (\ref{FPEfin})
because
\begin{align}
  \frac{\partial}{\partial {\bf r}}\phi(r)=\phi'(r)\frac{{\bf r}}{r}
\end{align}
At   the  same   time,   any   average  of   a   function  $F(r)$   is
time-independent. This is shown as follows
\begin{align}
  &\frac{d}{dt}\int d{\bf r}F(r)P({\bf r},t)
    \nonumber\\
  &=\int d{\bf r}F(r)
    \frac{1}{2}\frac{\partial}{\partial {\bf r}}\esc
    [{\bf r}]^T_\times\esc{\bf D}\esc [{\bf r}]_\times  \frac{\partial}{\partial {\bf r}}P({\bf r},t)  
    \nonumber\\
  &=- \frac{1}{2}\int d{\bf r}\frac{\partial F(r)}{\partial {\bf r}}\esc
    [{\bf r}]^T_\times\esc{\bf D}\esc [{\bf r}]_\times  \frac{\partial}{\partial {\bf r}}P({\bf r},t)  
    =0    \nonumber\\
  &=- \frac{1}{2}\int d{\bf r}F'(r)
    \underbrace{\frac{{\bf r}}{r}\esc
    [{\bf r}]^T_\times}_{=0}\esc{\bf D}\esc [{\bf r}]_\times  \frac{\partial}{\partial {\bf r}}P({\bf r},t)  
    =0
\end{align}
Now  consider  the probability  that  the  particle has  a  particular
modulus $r$. By definition it is given by the pushforward
\begin{align}
  P(r,t)=\int d{\bf r}\delta(r-|{\bf r}|)P({\bf r},t)
\end{align}
This  quantity  is, in  fact,  time-independent.  It takes  the  value
$P(r,0)$  that  it has  at  the  initial  time  and, also,  the  value
$P(r,\infty)=P(r,0)$ that has at equilibrium. Therefore, we have
\begin{align}
  P(r,0)&=\int d{\bf r}\delta(r-|{\bf r}|)P^{\rm eq}({\bf r})
          \nonumber\\
  &=\phi(r)\int d{\bf r}\delta(r-|{\bf r}|)=\phi(r)4\pi r^2
\end{align}
This means that the equilibrium solution must be given by
\begin{align}
  P^{\rm eq}({\bf r})=\frac{P(r,0)}{4\pi r^2}
\end{align}
Observe  that the  stationary solution  of  the FPE  depends on  the
initial distribution. For example, if the initial condition is  one in which
we are  certain that  the particle is  at a particular point  ${\bf r}_0$  on the
surface  of  the  unit  sphere,  i.e.  $|{\bf  r}_0|=1$,  the  initial
condition   is   $P({\bf   r},0)=\delta({\bf   r}-{\bf   r}_0)$.   The
corresponding  stationary  solution  of  the  FPE  with  this  initial
condition is
\begin{align}
  P^{\rm eq}({\bf r})=\frac{\delta(|{\bf r}|-1)}{4\pi}
  \label{Peq}
\end{align}

\subsection{Correlation function}
The equilibrium correlation  function of  the position of  a particle  perfoming a
Brownian motion on  the unit sphere can be  analytically computed, as follows. The
FPE (\ref{FPEfin}) can be written as
\begin{align}
  \partial_tP({\bf r},t)
  &={\cal L}
    P({\bf r},t)
\end{align}
where the Fokker-Planck operator is given by
\begin{align}
  \label{Lop}
  {\cal L}
  &=\frac{1}{2}\frac{\partial}{\partial {\bf r}}
    [{\bf r}]^T_\times\esc{\bf D}\esc [{\bf r}]_\times  \frac{\partial}{\partial {\bf r}}
\end{align}
The stationary correlation function can be expressed as \cite{riskenFokkerPlanckEquation1984}
\begin{align}
  \label{corr}
  \llangle{\bf r}_\alpha(t){\bf r}_\beta\rrangle =
  \int d{\bf r}P^{\rm eq}({\bf r}){\bf r}_\beta e^{{\cal L}t}{\bf r}_\alpha
\end{align}
where the exponential of the Fokker-Planck operator is defined in terms of its series expansion
\begin{align}
  e^{{\cal L}t}=\sum_{n=0}^\infty \frac{t^n}{n!}{\cal L}^n
\end{align}
We wish to compute the action  of the exponential operator on the unit
vector,  $e^{{\cal  L}t}{\bf  r}$.   The second  term  in  the  series
expansion is ${\cal  L}{\bf r}$ that we write  explicitly in component
form as
\begin{align}
  {\cal L}{\bf r}_\gamma
  &=-\frac{1}{2}\frac{\partial}{\partial {\bf r}_\mu}\epsilon_{\mu\alpha\mu'}{\bf r}_\alpha {\bf D}_{\mu'\nu'}\epsilon_{\nu'\beta\nu}{\bf r}_\beta\frac{\partial}{\partial {\bf r}_\nu}{\bf r}_\gamma
   \nonumber\\
  &=\frac{1}{2}{\bf D}_{\mu'\nu'}\epsilon_{\mu\alpha\mu'}\epsilon_{\mu\nu'\gamma} {\bf r}_\alpha 
    \nonumber\\
  &=\frac{1}{2}{\bf D}_{\mu'\nu'}\left(\delta_{\alpha\nu'}\delta_{\mu'\gamma}-\delta_{\alpha\gamma}\delta_{\mu'\nu'}\right)
    {\bf r}_\alpha 
    \nonumber\\
  &=\frac{1}{2}{\bf D}_{\gamma\alpha}
    {\bf r}_\alpha
    -\frac{1}{2}{\rm Tr}[{\bf D}] {\bf r}_\gamma
\end{align}
Therefore, 
\begin{align}
  {\cal L}{\bf r}=-{\bf A}\esc {\bf r}
\end{align}
where the  matrix ${\bf A}$  has been  defined in (\ref{A}).   It is
then obvious that
\begin{align}
  e^{{\cal L}t}{\bf r}=e^{-{\bf A}t}\esc{\bf r}
\end{align}
where the matrix exponential is introduced through the series expansion.

The correlation matrix (\ref{corr}) is now
\begin{align}
  \label{corr1}
  \llangle{\bf r}_\alpha(t){\bf r}_\beta\rrangle
  &=
    \int d{\bf r}P^{\rm eq}({\bf r}){\bf r}_\alpha\left[e^{-{\bf A}t}\right]_{\beta\beta'}{\bf r}_{\beta'}
    \nonumber\\
  &=
    \left[e^{-{\bf A}t}\right]_{\beta\beta'}   \int d{\bf r}P^{\rm eq}({\bf r}){\bf r}_\alpha{\bf r}_{\beta'}
    \nonumber\\
  &\overset{(\ref{Peq})}{=}
    \left[e^{-{\bf A}t}\right]_{\beta\beta'}   \frac{1}{3}\delta_{\alpha\beta'}
\end{align}
and we  conclude that the  equilibrium time correlation matrix  of the
position of the Brownian particle on the unit sphere is given by
\begin{align}
  \label{corrA}
  \llangle{\bf r}(t){\bf r}^T\rrangle=\frac{1}{3}e^{-{\bf A}t}
\end{align}

\subsection{Representation in terms of a rotation matrix}
We consider now a time-dependent rotation matrix $\boldsymbol{\cal R}$
\begin{align}
  \label{Rec}
  \boldsymbol{\cal R}
  &
    =\left(\begin{array}{ccc}{\bf c}_1,{\bf c}_2,{\bf c}_3\end{array}\right)         
\end{align}
where  ${\bf   c}_\alpha$  are   column  vectors.
Because $  \boldsymbol{\cal R}$ is  an orthogonal matrix,  its columns
${\bf c}_\alpha$ form  an orthonormal basis set. We propose  now a SDE
for the rotation matrix in such  a way that the basis vectors describe
a Brownian motion on the unit sphere.  In order to reach this goal, we
introduce   the  angular   velocity  $\boldsymbol{\omega}_0$   in  the
principal axis frame through
\begin{align}
\frac{d}{dt}{\boldsymbol{\cal R}}&\equiv-[\boldsymbol{\omega}_0]_\times\esc  \boldsymbol{\cal R}
                         \label{omega0}
\end{align}
to be compared with (\ref{omega}).  
Both angular velocities (\ref{omega}) and (\ref{omega0}) are related through
\begin{align}\label{omega0omega}
\boldsymbol{\omega}_0=\boldsymbol{\cal R}\esc\boldsymbol{\omega}  
\end{align}
We may use  (\ref{omega0}) as an \textit{inspiration}  to construct an
SDE in the space of rotation  matrices.  To achieve this objective, we
transform   the  angular   velocity  $\boldsymbol{\omega}_0$   into  a
stochastic process, this is
\begin{align}
\boldsymbol{\omega}_0 &\to {\bf C}\circ \frac{d{\bf W}_t}{dt}
\end{align}
and \textit{postulate} the  following Stratonovich  SDE for
the rotation matrix
\begin{align}
  \label{SDE-R}
  d{\boldsymbol{\cal R}}&=-\left[{\bf C}\circ d{\bf W}_t\right]_\times\esc  \boldsymbol{\cal R}
  &&\mbox{Stratonovich}\end{align}
In  terms of  the
columns ${\bf c}_\alpha$, we may write (\ref{SDE-R}) in the form
\begin{align}
  d{\bf c}_1&=-\left[{\bf C}\circ d{\bf W}_t\right]_\times\esc{\bf c}_1
              \nonumber\\
  d{\bf c}_2&=-\left[{\bf C}\circ d{\bf W}_t\right]_\times\esc{\bf c}_2
              \nonumber\\
  d{\bf c}_3&=-\left[{\bf C}\circ d{\bf W}_t\right]_\times\esc{\bf c}_3
              \label{SDE-c}
\end{align}
where in each equation we have  exactly the same value of the independent increment of the Wiener process $d{\bf W}_t$. Therefore, the
columns  of the  rotation  matrix experience  an anisotropic  Brownian
motion   on   the  surface   of   the   unit   sphere  of   the   form
(\ref{SDE-Str}). We conclude that  the postulated SDE (\ref{SDE-R})
for the rotation matrix is an alternative representation of anisotropic
Brownian motion.

Observe  that the  SDE (\ref{SDE-c})  has a  number of  conserved
quantities. By scalarly multiplying with ${\bf c}_\alpha$ we have
\begin{align}
  {\bf c}^T_\alpha\esc d{\bf c}_\beta
  &={\bf c}^T_\alpha\esc({\bf c}_\beta \times \esc{\bf C}\circ d{\bf W}_t)
    =({\bf c}_\alpha\times {\bf c}_\beta )^T \esc{\bf C}\circ d{\bf W}_t
\end{align}
and this implies
\begin{align}
&  d({\bf c}_\alpha^T\esc{\bf c}_\beta)
 =    {\bf c}^T_\alpha\esc d{\bf c}_\beta+  {\bf c}^T_\beta\esc d{\bf c}_\alpha
    \nonumber\\
  &=
  ({\bf c}_\alpha\times {\bf c}_\beta)^T  \esc{\bf C}\circ d{\bf W}_t
    +({\bf c}_\beta\times{\bf c}_\alpha)^T \esc{\bf C}\circ d{\bf W}_t
 =0
\end{align}
This means that both the modulus  of the vectors and the angle between
them    are    conserved    by     the    dynamics.    If    initially
${\bf   c}^T_\alpha\esc{\bf    c}_\beta=\delta_{\alpha\beta}$,   these
orthonormality conditions are mantained at all times, as expected.

For  future  reference, we  also  construct  the  Ito version  of  the
Stratonovich  SDE  \ref{SDE-c}).   By  analogy  with  (\ref{SDE-Str}),
(\ref{SDE-Ito}), the  Stratonovich SDE (\ref{SDE-c}) is  equivalent to
the following Ito SDE
\begin{align}
  d{\bf c}_\alpha
  &=-{\bf A}\esc{\bf c}_\alpha
    -[{\bf C}\esc d{\bf W}]_\times\esc{\bf c}_\alpha
\label{SDE-c-Ito}\end{align}
where ${\bf A}$ is given  by (\ref{A}).  Equation (\ref{SDE-c-Ito}) is
equivalent to the following Ito SDE for the rotation matrix
\begin{align}
  d\boldsymbol{\cal R}
  &=-{\bf A}\esc\boldsymbol{\cal R}
    -[{\bf C}\esc d{\bf W}]_\times\esc\boldsymbol{\cal R}
  &&\mbox{Ito}
\end{align}
This concludes our review of the anisotropic Brownian motion of a particle on the unit sphere.

\section{Appendix: Derivation of (\ref{SDELambda-S}) }
\label{App:2}We start from the Stratonovich SDE (\ref{SDELambda}) for the orientation
  \begin{align}
  \label{SDELambda-App}
  d\boldsymbol{\Lambda}
&=k_BT{\bf F}^{\rm th}dt +     {\bf B}^T\esc{\bf C}\esc d\tilde{\bf W}
\end{align}
where 
\begin{align}
{\bf C}&\equiv  (2k_BT)^{1/2}\boldsymbol{\cal D}_0^{1/2}
\end{align}
This is of the form of the Ito SDE (\ref{SDE-I}) with
\begin{align}
  {\bf A}&=k_BT{\bf F}^{\rm th}
           \nonumber\\
  \boldsymbol{\Theta}
  &={\bf B}^T\esc(2k_BT)^{1/2}\boldsymbol{\cal D}_0^{1/2}
\end{align}
The corresponidng Stratonovich SDE (\ref{SDE-S}) is now 
\begin{align}
  \label{dLdIto}
  d\boldsymbol{\Lambda}
&=\left(k_BT{\bf F}^{\rm th}-{\bf V}\right)dt +     {\bf B}^T\esc{\bf C}\esc\circ d\tilde{\bf W}
\end{align}
where the stochastic drift ${\bf V}$ is given  by (\ref{Vi}), which in the present case is
\begin{align}
  {\bf V}_i&=k_BT\left[{\bf B}\esc\boldsymbol{\cal R}^T\esc \boldsymbol{\cal D}_0^{1/2}\right]_{kj}\frac{\partial}{\partial \Lambda_k}
           \left[{\bf B}\esc\boldsymbol{\cal R}^T\esc \boldsymbol{\cal D}_0^{1/2}\right]_{ij}
             \nonumber\\
&=k_BT \left[{\bf B}\esc\boldsymbol{\cal R}^T\right]_{kk'}\left[\boldsymbol{\cal D}_0\right]_{k'i'}\frac{\partial}{\partial \Lambda_k}
  \left[{\bf B}\esc\boldsymbol{\cal R}^T\right]_{ii'}
  \label{Vi-1}
\end{align}
The rotation matrix and the kinematic operator satisfy
\begin{align}
    {\bf B}\esc\boldsymbol{\cal R}^T={\bf B}^{T}
  \label{BRB}
\end{align}
as can  be shown  from the  results in Sec.   K.1 of  the Supplemental
Material  of   Ref.   \cite{espanol2024}.    By  using   the  property
(\ref{BRB}), (\ref{Vi-1}) becomes
\begin{align}
  {\bf V}_i
&=k_BT \left[\boldsymbol{\cal D}_0\right]_{k'i'}{\bf B}_{k'k}\frac{\partial}{\partial \Lambda_k}
  {\bf B}_{i'i}\overset{(\ref{eq:366})}{=}k_BT{\bf F}^{\rm th}_i
\end{align}
Therefore, (\ref{dLdIto}) becomes
\begin{align}
    d\boldsymbol{\Lambda}
&=     {\bf B}^T\esc{\bf C}\esc\circ d\tilde{\bf W}
\end{align}
which is (\ref{SDELambda-S}), as we wished to demonstrate.
\section{Appendix: Derivation of (\ref{R-sphere2}) }
\label{App:3}
By using the chain rule of ordinary calculus with the Stratonovich SDE
(\ref{SDELambda-S})  we wish  to  obtain a  Stratonovich  SDE for  the
rotation matrix. Indeed
\begin{align}
  d\boldsymbol{\cal R}
  &=\frac{\partial \boldsymbol{\cal R}}{\partial \boldsymbol{\Lambda}}\esc d\boldsymbol{\Lambda}
\overset{(\ref{SDELambda-S})}{=}\frac{\partial \boldsymbol{\cal R}}{\partial \boldsymbol{\Lambda}}\esc{\bf B}^T\esc{\bf C}\esc \circ d\tilde{\bf W}
      \label{R-sphere3}
\end{align}
From the  results in Sec.   K.1 of  the Supplemental Material  of Ref.
\cite{espanol2024}, the derivative of the rotation matrix with respect
to the orientation is given by
  \begin{align}
    \label{dRdL}
    \frac{\partial \boldsymbol{\cal R}_{\mu\nu}}{\partial\boldsymbol{\Lambda}_\alpha}
  &=-{\bf B}^{-1}_{\alpha\beta}\epsilon_{\mu\beta\delta}\boldsymbol{\cal R}_{\delta\nu}
\end{align}
Using this derivative in (\ref{R-sphere3}) leads to 
\begin{align}
  \label{r1}
d\boldsymbol{\cal R}_{\mu\nu}&=  \frac{\partial \boldsymbol{\cal R}_{\mu\nu}}{\partial \boldsymbol{\Lambda}_\alpha}{\bf B}_{\alpha\alpha'}
\boldsymbol{\cal R}^T_{\alpha'\beta}[{\bf C}\esc d\tilde{\bf W}_t]_\beta
                               \nonumber\\
&=-  {\bf B}^{-1}_{\alpha\beta}\epsilon_{\mu\beta\delta}\boldsymbol{\cal R}_{\delta\nu}
  {\bf B}_{\alpha\alpha'}
\boldsymbol{\cal R}^T_{\alpha'\beta'}[{\bf C}\esc d\tilde{\bf W}_t]_{\beta'}
  \nonumber\\
&=-\epsilon_{\mu\beta\delta}\boldsymbol{\cal R}_{\delta\nu}
\underbrace{  {\bf B}^{-1}_{\alpha\beta}  {\bf B}_{\alpha\alpha'}
\boldsymbol{\cal R}^T_{\alpha'\beta'}}_{[{\bf B}^{-T}\cdot{\bf B}\cdot\boldsymbol{\cal R}^T]_{\beta\beta'}}[{\bf C}\esc d\tilde{\bf W}_t]_{\beta'}
\end{align}
The property (\ref{BRB}) implies 
\begin{align}
  {\bf B}^{-T}\esc{\bf B}\esc\boldsymbol{\cal R}^T=\mathbb{1}
\end{align}
and, therefore,
\begin{align}
  d\boldsymbol{\cal R}_{\mu\nu}
  &=-  \epsilon_{\mu\beta\delta}\boldsymbol{\cal R}_{\delta\nu}
    \underbrace{  {\bf B}^{-1}_{\alpha\beta}  {\bf B}_{\alpha\alpha'}
    \boldsymbol{\cal R}^T_{\alpha'\beta'}}_{\delta_{\beta\beta'}}[{\bf C}\esc d\tilde{\bf W}_t]_{\beta'}
    \nonumber\\
  &=      -\epsilon_{\mu\alpha\beta}\boldsymbol{\cal R}_{\beta\nu}\left[{\bf C}\esc d{\bf W}_t\right]_\alpha 
\end{align}
which is the desired result (\ref{R-sphere2}).

\end{document}